\documentclass[preprint]{aastex62}
\usepackage{natbib}
\usepackage{graphicx}
\usepackage{epstopdf}
\usepackage{mathtools}
\usepackage{bm}
\newcommand{\be}{\begin{equation}}
\newcommand{\ee}{\end{equation}}
\newcommand{\ba}{\begin{eqnarray}}
\newcommand{\ea}{\end{eqnarray}}
\newcommand{\bnabla}{\mbox{\boldmath$\nabla$}}
\newcommand{\nn}{\mbox{} \nonumber \\ \mbox{}}

\begin{document}
\title{QED Phenomena in an Ultrastrong Magnetic Field. I.\\ Electron-Photon Scattering, Pair Creation and Annihilation}

\author{Alexander Kostenko}
\affil{Department of Astronomy and Astrophysics, University of Toronto, 50 St. George Street, Toronto, ON M5S 3H4, Canada}
\author{Christopher Thompson}
\affil{Canadian Institute for Theoretical Astrophysics, 60 St. George Street, Toronto, ON M5S 3H8, Canada}

\received{2018 July 19}
%\accepted{2018 September 10}
\published{2018 December 10; Erratum added}
%\submitjournal{ApJ}
\shorttitle{QED in an Ultrastrong Magnetic Field I.}
\shortauthors{Kostenko \& Thompson}

%\date{\today}

\begin{abstract}
We evaluate several basic electrodynamic processes as modified by the presence of a very strong magnetic field, exceeding $B_{\rm Q} \equiv m^2/e = 4.4\times 10^{13}$ G. These results are needed to build models of dissipative phenomena outside magnetars and some other neutron stars. Differential and total cross sections and rates are presented for electron-photon scattering, the annihilation of an electron-positron pair into two photons, the inverse process of two-photon pair creation, and single-photon pair creation into the lowest Landau state. The relative importance of these interactions changes as the background magnetic field grows in strength. The particle phase space relevant for a given process may be restricted by single-photon pair creation, which also opens up efficient channels for pair multiplication, e.g. in combination with scattering.  Our results are presented in the form of compact formulae that allow for relativistic electron (positron) motion, in the regime where Landau excitations can be neglected (corresponding to $10^3B_{\rm Q} \gg B \gg B_{\rm Q}$ for moderately relativistic motion along the magnetic field). Where a direct comparison is possible, our results are tested against earlier calculations, and a brief astrophysical context is provided.  A companion paper considers electron-positron
scattering, scattering of electrons and positrons by ions, and relativistic electron-ion bremsstrahlung.
\end{abstract}

\section{Introduction}

The electromagnetic interactions of electrons and positrons are strongly modified, and frequently complicated, by the
presence of an ultrastrong magnetic field \citep{HL2006}.  Considerable attention has previously been given to the resonant scattering of photons by $e^\pm$ (see \citealt{MP1983B, BAM1986, DH1989, DH1991, Gonthier2000}) in the magnetospheres of
pulsars and magnetars,  involving Landau excitations in intermediate states.  But close to the surface of a magnetar, the magnetic field is so strong that interacting electrons and positrons may be substantially confined to the lowest Landau state. The hard X-ray emission of magnetars, which rises to at least 100 keV \citep{Kuiper2006, Mereghetti2006}, may probe a plasma state consisting of a relatively dense gas of trans-relativistic electrons, positrons, and ions near the neutron star surface \citep{TB2005}. Related physical conditions are encountered in pair-loaded
coronae around bursting magnetars \citep{TD2001} and in the magnetospheres of inspiraling and colliding neutron stars \citep{HL2001}. 

Near the magnetar surface, electron-photon scattering is nonresonant but still substantially anisotropic, and quantum recoil effects are important.  Bremsstrahlung becomes an important source of hard X-rays, and the behavior of electrons and positrons is modified by both scattering off ions in the upper atmosphere of the neutron star and mutual electron-positron scattering.  The annihilation of electrons and positrons is suppressed as the magnetic field $B$ rises above $B_{\rm Q} \equiv m^2/e$ \citep{DB1980}, but the cross section for photon collisions into $e^\pm$ pairs is 
enhanced compared with an unmagnetized vacuum \citep{KM1986}.

The preceding literature provides incomplete coverage of these quantum electrodynamic (QED) processes. Consider, for
example, electron-photon (Compton) scattering.  Classical and quantum descriptions of this process in a strong magnetic
field were worked out some time ago: see \citet{Canuto1971}, \citet{Ventura1979}, and \citet{DH1986}, \citet{DH1991},
respectively.  The full
QED description involves a complicated sum over intermediate-state  Landau levels.  A more compact result for the cross section is of value in concrete (e.g. Monte Carlo) calculations and, as we show, is well motivated when (i) the energy of
Landau excitations is much larger than the electron rest mass $m$, corresponding to $B \gg B_{\rm Q}$,  but (ii)
nonlinearities due to vacuum polarization are still weak ($B \ll 10 \alpha_{\rm em}^{-1} B_{\rm Q} \sim 10^3 B_{\rm Q}$;
\citealt{HL2006}).  

A significant benefit of such a compact formulation is that additional effects, such as the conversion of a scattered or
emitted photon into an electron-positron pair, are much easier to analyze.  These have a rich behavior and, as we show,
have significant impact on the net rates of electron-photon scattering or pair annihilation.  Their analysis forms a
significant part of this paper. 

We first provide a brief overview of the QED rules in a background magnetic field that will allow the reader to follow our calculations (Section \ref{s:qed}).  We adopt the same basic procedure for each process considered:  the QED calculation is
performed from scratch, using the appropriate magnetized electron/positron wave functions \citep{ST1968, MP1983},  with
incoming, outgoing, and intermediate $e^\pm$ lines being restricted to the lowest Landau level.  In some cases,
such as electron-photon scattering, the full Landau level sum had previously been worked out and our
result was found to be consistent with the truncation of the prior calculation (Section \ref{Compton}).  In other cases,
the more complete calculation is sufficiently complex to have inhibited previous attempts at calculations, especially those
involving relativistic $e^\pm$ motion.

A variety of pair creation processes operate in the magnetar magnetosphere. In a background magnetic field, a single photon
is kinematically capable of creating a pair \citep{Erber1966, LLT4}.   We derive the rate of single-photon
pair creation in the regime where the created pair is restricted to the lowest Landau level, using a detailed balance
argument (Section \ref{1gamma}).  Then we evaluate the rate at which electron scattering mediates the conversion of an
energetic photon to a pair (Section \ref{s:assist}).
This rate is greatly enhanced by a pole involving the annihilation of the incoming electron with a virtual positron.
An accurate analytic approximation to the cross section is derived by integrating over this pole.
Then in Section \ref{2gammareview} we reconsider the collision of two gamma rays into an $e^\pm$ pair 
\citep{KM1986, CT2008}.  The kinematic constraints on this process are weakened in the presence of a strong
background magnetic field:  a soft photon is now
able to trigger the conversion of a harder photon with an energy only slightly smaller than $2m$.  A simple form for the
collision cross section is written down in terms of invariants of the photon momenta.

Finally, we analyze two-photon pair annihilation in Section \ref{s:ann2g}, which is suppressed in an ultrastrong magnetic
field by the concentration of the electron wave functions transverse to the magnetic field \citep{DB1980}.  The cross
section for annihilation is related to the cross section for two-photon pair creation by an integral formula.  We
quantify how annihilation is restricted by the reconversion of one (or both) final state photons
to a pair.  The net annihilation rate is sharply suppressed as the annihilating pair becomes mildly relativistic.  By the
same token, the annihilation of an electron and positron into a single photon -- which is allowed in the presence of the
background magnetic field \citep{DB1980} -- is almost completely suppressed by the rapid reconversion of
the photon to a pair.

Our results are summarized in Section \ref{s:concl}.
A companion paper \citep{KT2018} evaluates several other relevant QED processes in a strong magnetic field, including
electron-positron scattering, $e^\pm$-ion scattering, and relativistic $e^\pm$-ion bremsstrahlung.

We adopt natural units ($\hbar = c = k_B = 1$) throughout this paper, along with the $(+---)$ metric signature.  The
Dirac gamma matrix convention is the same as that used by \citet{MP1983},
\be
\gamma^0 = \begin{pmatrix}
1 & 0 \\ 
0 & -1 
\end{pmatrix};\quad\quad
\gamma^j = \begin{pmatrix}
0 & \sigma^j  \\ 
-\sigma^j  & 0 
\end{pmatrix}, 
\ee
where each $0$ and $1$ element denotes a $2\times 2$ matrix, and $\sigma^j$ are the usual Pauli matrices. Landau
gauge ${\bm A} = Bx \hat y$ is chosen for the background vector potential, and we alternatively use
Cartesian coordinates $(x, y, z)$ and spherical coordinates $(\theta,\phi)$ (with the axis $\theta = 0$
aligned with $\hat z$) to describe the wavevectors of interacting particles.

\section{QED Interactions in a Background Magnetic Field}\label{s:qed}

We now review some basic properties of electron/positron and photon states in a background magnetic field,
in preparation for our evaluation of various cross sections.  The main choice to be made is of the
electron/positron wave function, in which we follow \citet{ST1968} and \citet{MP1983}.  Then the wave function of an electron
moving along the magnetic field is connected by a simple Lorentz transformation to the wave function
of an electron at rest.  Two considerations lead us to limit the strength of the background magnetic field to 
$\lesssim 10^3\,B_{\rm Q}$:  (i) vacuum polarization significantly modifies the photon dispersion relation in stronger
magnetic fields \citep{Adler1971}, and (ii) the decay rate of a photon of energy $\omega > 2m$ becomes of the order of $\omega$, 
so that a propagating photon state loses meaning.

\subsection{Photons}\label{s:photon}

The two polarization states of photons of frequency $\omega \ll |e|B/m = (B/B_{\rm Q})m$ show a strong asymmetry in
their scattering and emission cross sections \citep{MV1979, HL2006}.   The ordinary (O) mode interacts much
more strongly with electrons than the extraordinary (E) mode, because a significant component of its electric vector
is directed along the background magnetic field.   The O-mode cross sections for electron scattering and bremsstrahlung
emission are comparable in magnitude to the unmagnetized values.   This asymmetry 
disappears in a narrow range of propagation directions about the magnetic axis; it also disappears at a critical electron
density where the contributions of vacuum polarization and plasma to the dielectric tensor nearly cancel \citep{HL2006}.

The photon wave function is normalized as
\be
A^\mu(x^\nu) = {\varepsilon^\mu\over (2\omega L^3)^{1/2}} e^{-ik\cdot x};  \quad\quad  k^\mu = \omega(1, \hat k).
\ee
Here, $k^\mu$ and $\varepsilon^\mu$ are the wavevector and polarization 4-vectors, and $L^3$ is a normalization volume.
Excepting near a vacuum-plasma resonance, both polarization modes
are highly elliptically polarized.   A photon propagating in the direction $\hat k = (\hat k_x, \hat k_y, \hat k_z) = 
(\sin\theta\cos\phi, \sin\theta\sin\phi, \cos\theta)$ has a unit electric vector parallel to 
$\hat k \times \hat B$ in the E-mode, and parallel to $\hat k\times (\hat k\times\hat B)$ in the O-mode, i.e., 
\be\label{eq:pol}
\varepsilon_{\rm O}^z = \sin\theta; \quad \varepsilon_{\rm O}^\pm = \varepsilon_{\rm O}^x \pm i\varepsilon_{\rm O}^y = 
-\cos\theta e^{\pm i\phi}.
\ee
For the hard X-rays and gamma rays of interest here, vacuum polarization is the dominant correction
to the dielectric response, and the expressions given for $\varepsilon^i_{\rm O,E}$ are essentially exact:  they are 
accurate to $O(4\pi n_\pm m_e c^2/B^2)$, where $n_\pm$ is the number density of electrons and positrons \citep{MV1979}.

When Landau resonances are kinematically forbidden, the polarization dependence of the processes we consider
reduces to a dependence on $\varepsilon^z$.  This effectively decouples the E-mode:
\be
\varepsilon_{\rm E}^z = 0;  \quad \varepsilon_{\rm E}^\pm = \mp ie^{\pm i\phi}.
\ee
The coupling to the E-mode is restored when virtual Landau excitations are included in a matrix element.  This 
introduces terms in each matrix element involving the $\varepsilon^\pm$ polarization components, but with a magnitude
suppressed by $\sim m \omega/e B$ at frequencies well below the first Landau resonance.   This means that the
nonresonant E-mode cross section is generally suppressed by a factor $\sim (m \omega/e B)^2$ compared with that 
of the O-mode.

\subsection{Electrons and Positrons}\label{s:pairs}

Quantum states of an electron or positron of charge $q = \mp e$ in a magnetic field ${\bm B} = \bnabla\times{\bm A} 
= B\hat z$ are characterized by two conserved components of the generalized momentum:  a longitudinal momentum $p_z$,
and a gauge-dependent transverse momentum $q{\bm A}$ marking a center of gyration ${\bm x}_\pm$ in the plane
perpendicular to ${\bm B}$.  In the Landau gauge ${\bm A} = Bx\hat y$, the electron wave function is localized
in coordinate $x$, within a strip of width $\sim \lambda_B \equiv (|e|B)^{-1/2} = (B/B_{\rm Q})^{-1/2} m^{-1}$.

This means, for example, that an electron which absorbs momentum $-\Delta k_y$ by scattering a photon will see its
center of gyration shift by $\Delta x_- = +\Delta k_y/|e|B = \lambda_B^2\,\Delta k_y$; and that an annihilating pair of
positive and negative electrons whose centers of gyration $x_\pm$ are displaced relative to each other will emit photon(s) carrying net $y$-momentum $\Delta k_y = |e|B(x_+-x_-)$.  On the other hand, in this particular gauge there is no
conserved $x$-momentum, meaning that the $x$-momentum carried by photons in the final state is constrained only by the
conservation of energy.  The total cross section and the kinematic constraints on it are of course always independent 
of this gauge choice\footnote{In a classical approximation,
the translational invariance of the magnetic field implies the conservation of transverse canonical momentum 
${\bm p} + q{\bm A}$, and one recovers the Lorentz force $d{\bm p}/dt = -q\left[\partial{\bm A}/\partial t + 
({\bm v}\cdot\bnabla){\bm A}\right] = q{\bm v}\times {\bm B}$.}.

The energy levels of the electron or positron are \citep{LLT4}
\be\label{eq:en}
E^2 = p_z^2 + m^2 + |q|B(2l+1) - qB\sigma,
\ee
where $l \geq 0$ is an integer labeling the orbital angular momentum of the mode, and $\sigma$ is the 
eigenvalue of the spin operator:
\be
\Sigma_z = \begin{pmatrix}\sigma_z & 0\\0 & \sigma_z \end{pmatrix}
\ee
as evaluated in the particle rest frame ($p_z = 0$).  Here, $\sigma_z$ is the $2\times 2$ dimension Pauli matrix.  
In the lowest Landau state ($E = m$), the electron has spin $\sigma = -1$ and the 
positron $\sigma = +1$, as expected from the nonrelativistic expansion of the Dirac equation.  These spin labels
can be continuously extended to finite $p_z$, as described in Section \ref{s:spinors}.

The localization of the electron wave function transverse to the magnetic field depends on the sum of the
last two terms on the right-hand side of Equation (\ref{eq:en}), and so we adopt the simplified notation
\be
E^2 = p_z^2 + m^2 + p_n^2;   \quad\quad   p_n^2 \equiv 2n|e|B \equiv E_{n0}^2 - m^2,
\ee
where $n = l + {1\over 2}[1-\sigma\cdot{\rm sgn}(q)]$.

In what follows, we assume that the initial electron or positron sits in the lowest Landau state, given
the short timescale for radiative de-excitation from $n > 0$.  Particles in all processes are also assumed to
carry a small enough kinetic energy to prevent excitations to $n > 0$ in the final state, or in resonances.

\subsection{Dirac Spinors}\label{s:spinors}

The electron/positron wave functions are written, following \citet{ST1968} and \citet{MP1983}, as
\be
\left [ {\psi}_\mp^{(\sigma)}(x^\mu) \right ]_{p_z,n,a} = 
\left\{\begin{matrix} 
e^{-ip\cdot x}\,u_{n,a}^{(\sigma)}({\bm x}) \quad\quad (\rm electrons);\\ 
e^{ip\cdot x}\,v_{n,a}^{(\sigma)}({\bm x}) \quad\quad (\rm positrons). \\ 
\end{matrix}\right. 
\ee
Here, $\sigma = \pm 1$ labels the spin state, $a$ the center of gyration, and $p^\mu$ the momentum 4-vector,
\be
p^\mu = (E, 0, p^y, p^z);  \quad\quad p_y = a qB = {\rm sgn}(q){a\over \lambda_B^2}.
\ee
Under charge conjugation, the sign of $p^\mu$ reverses, and so the gyration center remains fixed.

The choice of the positive- and negative-energy spinors $u_{n,a}^{(\sigma)}({\bm x})$, 
$v_{n,a}^{(\sigma)}({\bm x})$, is guided by the requirement that for finite $p_z$ they 
be continuously related to the spinor of a particle at rest.   Their general form is \citep{JL1949}
\be
\begin{bmatrix}
C_1\phi_{n-1}(x) \\ C_2 \phi_{n}(x) \\ C_3 \phi_{n-1}(x) \\ C_4 \phi_{n}(x)
\end{bmatrix}, 
\ee
where the $\phi_n$ are harmonic oscillator wave functions,
\be
\phi_n(x-a) = \frac{1}{L(\pi^{\frac{1}{2}} \lambda_B 2^nn!)^{\frac{1}{2}}} H_n\left( \frac{x-a}{\lambda_B} \right)
e^{-(x-a)^2/2\lambda_B^2},
\ee
and $H_n$ is the $n^{\rm th}$-order Hermite polynomial.  For a particle at rest ($p_z = 0$),
\be\label{eq:sp}
\begin{bmatrix}  C_1 \\ C_1 \\ C_3 \\C_4 \end{bmatrix} = 
\frac{1}{\sqrt{2\epsilon E_{n0}(\epsilon E_{n0} + m)}}
\left[ \begin{bmatrix} \epsilon E_{n0} + m \\ 0 \\ 0 \\ i p_n \end{bmatrix} D_{\sigma = +1} + 
\begin{bmatrix} 0 \\ \epsilon E_{n0} + m \\ -i p_n \\ 0 \end{bmatrix} D_{\sigma = -1} \right].
\ee
Here, $\epsilon = +1 (-1)$ corresponds to positive (negative) energy states.
It is easy to check that these spinors are eigenstates of the $z$-component of spin,
\be
\int d^3x \bar{\psi} \Sigma_z \psi = D_{\sigma=+1}^2 - D_{\sigma=-1}^2 = \pm 1.
\ee
Hence, $D_{\sigma=+1} = 1$ ($D_{\sigma=-1} = 1$) corresponds to a state of spin-up (spin-down).  In the lowest
Landau state $n=0$, only $D_{\sigma = -1} = 1$ is available for the positive-energy state (the function 
$\phi_n$ is undefined for $n = -1$).

To obtain the spinors at finite $p_z$, one applies the Lorentz boost parallel to ${\bm B}$,
\be\label{eq:psiboost}
\psi \rightarrow \exp\left( -\frac{1}{2} \alpha_z \rho \right) \psi ; \quad\quad 
\alpha_z = \begin{pmatrix} 0 & \sigma_z \\ \sigma_z & 0 \end{pmatrix}.
\ee
Here, $\beta \equiv \tanh(\rho)$ is the speed of the boost, so that $\gamma = \cosh(\rho)$. 
Taking into account that $\alpha_z^2 = I$,  we have
\be
\exp\left( -\frac{1}{2} \alpha_z \rho \right) = 
I \cosh\left( \frac{1}{2}\rho \right) - \alpha_z \sinh\left( \frac{1}{2} \rho \right).
\ee

Applying this transformation to the spinors in Equation (\ref{eq:sp}), and dividing by $\sqrt{\gamma}$ to 
compensate for the longitudinal contraction of the wave packet under the boost, one obtains
\be\label{eq:spinors}
u_{n,a}^{(-1)}({\bm x})=\frac{1}{f_n}\begin{bmatrix}
-ip_zp_n\phi_{n-1}\\
(E+E_{0n}) (E_{0n}+m)\phi_n
\\ -ip_n(E+E_{0n})\phi_{n-1}
\\ -p_z(E_{0n}+m)\phi_{n}
\end{bmatrix} ; \quad\quad
u_{n,a}^{(+1)}({\bm x})=\frac{1}{f_n}\begin{bmatrix}
(E+E_{0n}) (E_{0n}+m)\phi_{n-1}\\
-ip_z p_n\phi_n
\\ p_z(E_{0n}+m)\phi_{n-1}
\\ ip_n (E+E_{0n}) \phi_n
\end{bmatrix}
\ee
for the positive-energy spinors and 
\be
v_{n,a}^{(+1)}({\bm x})=\frac{1}{f_n}\begin{bmatrix}
-p_n (E+E_{0n})\phi_{n-1}\\
-ip_z(E_{0n}+m)\phi_n \\
-p_z p_n\phi_{n-1} \\
i(E+E_{0n}) (E_{0n}+m)\phi_n
\end{bmatrix} ; \quad\quad
v_{n,a}^{(-1)}({\bm x})=\frac{1}{f_n}\begin{bmatrix}
-ip_z(E_{0n}+m)\phi_{n-1} \\
-p_n (E+E_{0n}) \phi_n \\
-i(E+E_{0n}) (E_{0n}+m)\phi_{n-1} \\
p_z p_n\phi_{n} \\
\end{bmatrix} 
\ee
for the negative-energy spinors.  Here, we introduce $f_n = 2L\sqrt{E E_{0n}(E_{0n} + m)(E_{0n} + E)}$.
It is straightforward to check that $u^{(\pm 1)}_{n,a}$ and $v^{(\mp 1)}_{n,a}$ are connected by
exchanging $\sigma, p_z, E, E_{0n} \rightarrow -\sigma, -p_z, -E, -E_{0n}$.

The wave functions derived by \citet{JL1949} do not have this property of being continuously related
by a Lorentz transformation to the wave function of an electron/positron with vanishing $p_z$.  Wave functions
equivalent to ours (up to a trivial phase factor) can alternatively be derived by requiring them to be
eigenstates of an appropriately defined magnetic moment operator \citep{ST1968, MP1983}.

All the cross sections evaluated in this paper are independent of the choice of spinor basis, since they are
effectively summed (averaged) over the single admissible spin state of each outgoing (ingoing) electron or 
positron \citep{MP1983}.

\subsection{Rules for Calculating Matrix Elements}\label{s:rules}

We complete our review of QED amplitudes in strong magnetic fields by summarizing the Feynman rules as expressed in 
coordinate space and some computational procedures.

1. A vertex between photon and electron lines is written as the integral
\ba
  && -ie\int d^4x \left[ \bar{\psi}_-^{(\sigma_I)}(x) \right ]_{p_{z,I}, n_I, a_I} \gamma_\mu A^\mu(x) 
\left [\psi_-^{(\sigma_i)}(x) \right]_{p_{z,i}, n_i, a_i} \nn
        &&\quad\quad  = -{ie\over (2\omega L^3)^{1/2}} \int d^4x \,e^{-i(p_i \pm k - p_I)\cdot x} \,
      \bar u_{n_I,a_I}^{(\sigma_I)}({\bm x}) \gamma_\mu \varepsilon^\mu  u^{(\sigma_i)}_{n_i,a_i}({\bm x}).
\ea
Here, $i$ and $I$ label incoming and internal positive-energy electron states, respectively, and the photon is either
absorbed (wavevector $+k^\mu$) or emitted ($-k^\mu$).  The vertex between an incoming electron
and an internal positron is obtained by substituting $-p_I$ and $\bar v^{(-\sigma_I)}_{n_I,a_I}$ for 
$p_I$ and $\bar u^{(\sigma_I)}_{n_I,a_I}$.  

2. An internal electron line is represented by the propagator in coordinate space, 
\ba\label{ComptProp}
G_F(x'-x) &=& -i\int L\frac{da_I}{2\pi \lambda_B^2}\int L\frac{dp_{z,I}}{2 \pi}
\sum_{n_I=0}^{\infty} \biggl[ \theta(t'-t)\sum_{\sigma_I} u_{n_I,a_I}^{(\sigma_I)}({\bm x}')
\bar{u}_{n_I,a_I}^{(\sigma_I)}({\bm x})e^{-iE_I(t'-t)}e^{i{\bm p}_I\cdot({\bm x}'-{\bm x})} \nn
&& \quad\quad - \theta(t-t')\sum_{\sigma_I} 
v_{n_I,a_I}^{(\sigma_I)}({\bm x}')\bar{v}_{n_I,a_I}^{(\sigma_I)}({\bm x})
e^{iE_I (t'-t)}e^{-i{\bm p}_I\cdot({\bm x}'-{\bm x})} \biggr].
\ea

3. The combined integral over $t$ and $t'$ generates a combination of an energy delta function and an energy
denominator:
\ba
\mp i\int dt \int dt' \theta[\mp(t-t')]\,e^{i(E_f + \omega_f \mp E_I)t'}\, e^{-i(E_i + \omega_i \mp E_I)t}
= \frac{2\pi \delta(E_i + \omega_i - E_f - \omega_f)}{E_i + \omega_i \mp E_I}.
\ea
Here, $i$ and $f$ label incoming and outgoing particles.  The contribution of an excited Landau state to 
a given term in the matrix element is suppressed by a factor $E_I^{-1} \simeq (2n|e|B)^{-1/2}$ away from resonance.  
However, the suppression of the net rate is generally stronger as the result of a cancellation between
$S_{fi}[1]$ and $S_{fi}[2]$.  

4. The contraction of the electric polarization vector with $\gamma$ matrices is
\be\label{eq:amatrix}
\gamma_0\gamma_{\mu}\varepsilon_i^\mu = -\begin{pmatrix}
0 & 0 & \varepsilon_i^z  &\varepsilon_i^- \\ 
0 &0  & \varepsilon_i^+ & -\varepsilon_i^z \\ 
\varepsilon_i^z & \varepsilon_i^- & 0 &0 \\ 
\varepsilon_i^+ & -\varepsilon_i^z &0  &0 
\end{pmatrix};  \quad\quad  (i = {\rm O}, {\rm E}). 
\ee 

5. The matrix element $S_{fi}$ includes energy and momentum delta functions that, once squared, are handled according to
(e.g. in the case of Compton scattering)
\begin{gather*}
\left[2\pi \delta(p_{z,i} + k_{z,i} - p_{z,f} - k_{z,f})\right]^2 \;\rightarrow\;  
 L (2\pi) \delta(p_{z,i} + k_{z,i} - p_{z,f} - k_{z,f}); \\
\left[2\pi \delta\left(k_{y,i} - k_{y,f} - \frac{a_i-a_f}{\lambda_B^2}\right) \right]^2 \;\rightarrow\;  
L (2\pi) \delta\left(k_{y,i} - k_{y,f} - \frac{a_i-a_f}{\lambda_B^2}\right); \\
\left[2\pi \delta(E_i + \omega_i - E_f - \omega_f)\right]^2 \;\rightarrow\;  
T (2\pi) \delta(E_i + \omega_i - E_f - \omega_f).
\end{gather*}
Here, $T$ is the normalization time.  

The delta function in $p_y$ has a term from the change in the guiding center $a$ of the scattering charge.
No delta function in $p_x$ appears for our choice of background gauge.  For the sake of brevity, such a 
combination of delta functions will be written in the following way:
\be
\delta^{(3)}_{fi}(E,p_y,p_z).
\ee

6. Summing over the phase space of a final-state photon involves the integral
\be 
\int L^3 {\omega_f^2 d\omega_f d\Omega_f\over (2\pi)^3}, 
\ee
where $\Omega_f$ is solid angle.  For a final-state electron or positron, there is no sum over the $x$-component of
momentum, hence the integral
\be
{|e|B\over 2\pi} \int L da_f  \int L {dp_{z,f}\over 2\pi} = \int L{da_f\over 2\pi\lambda_B^2} \int L{dp_{z,f}\over 2\pi}.
\ee

\section{Electron-Photon Scattering} \label{Compton}

\begin{figure}[h]
%\includegraphics[scale=0.55]{NonResLabelPDF.pdf}
%\plotone{NonResLabelPDF.pdf}
\plotone{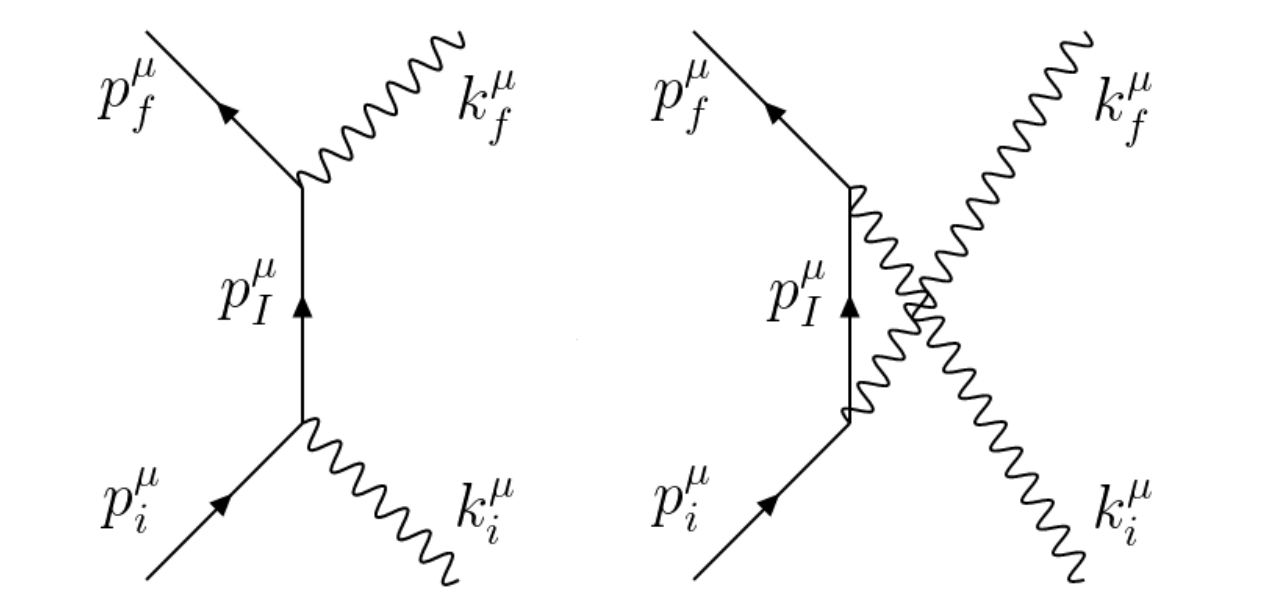}
\caption{Feynman diagrams for Compton scattering.}
\label{NR}
\end{figure}

We first consider nonresonant electron-photon scattering, $e^\pm + \gamma \rightarrow e^\pm + \gamma$, as modified by a
strong magnetic field (Figure \ref{NR}).  Previous QED calculations  
\citep{Herold1979, MP1983B, BAM1986, DH1986, DH1991, Gonthier2000, Baring2005} have focused on the situation where the initial photon is energetic enough to
excite the scattering charge to a higher Landau level.  We consider the case where Landau resonances are kinematically
forbidden even for initial photon energies around $m$.  We demonstrate a good agreement between a truncated formula for
the nonresonant scattering cross section and the full QED result.   In this situation, the initial photon energy
is restricted by single-photon pair creation;  photons approaching this pair creation threshold have an enhanced 
scattering cross section.

\begin{figure} [h]
\epsscale{0.75}
%\plotone{domegaPDF.pdf}
\plotone{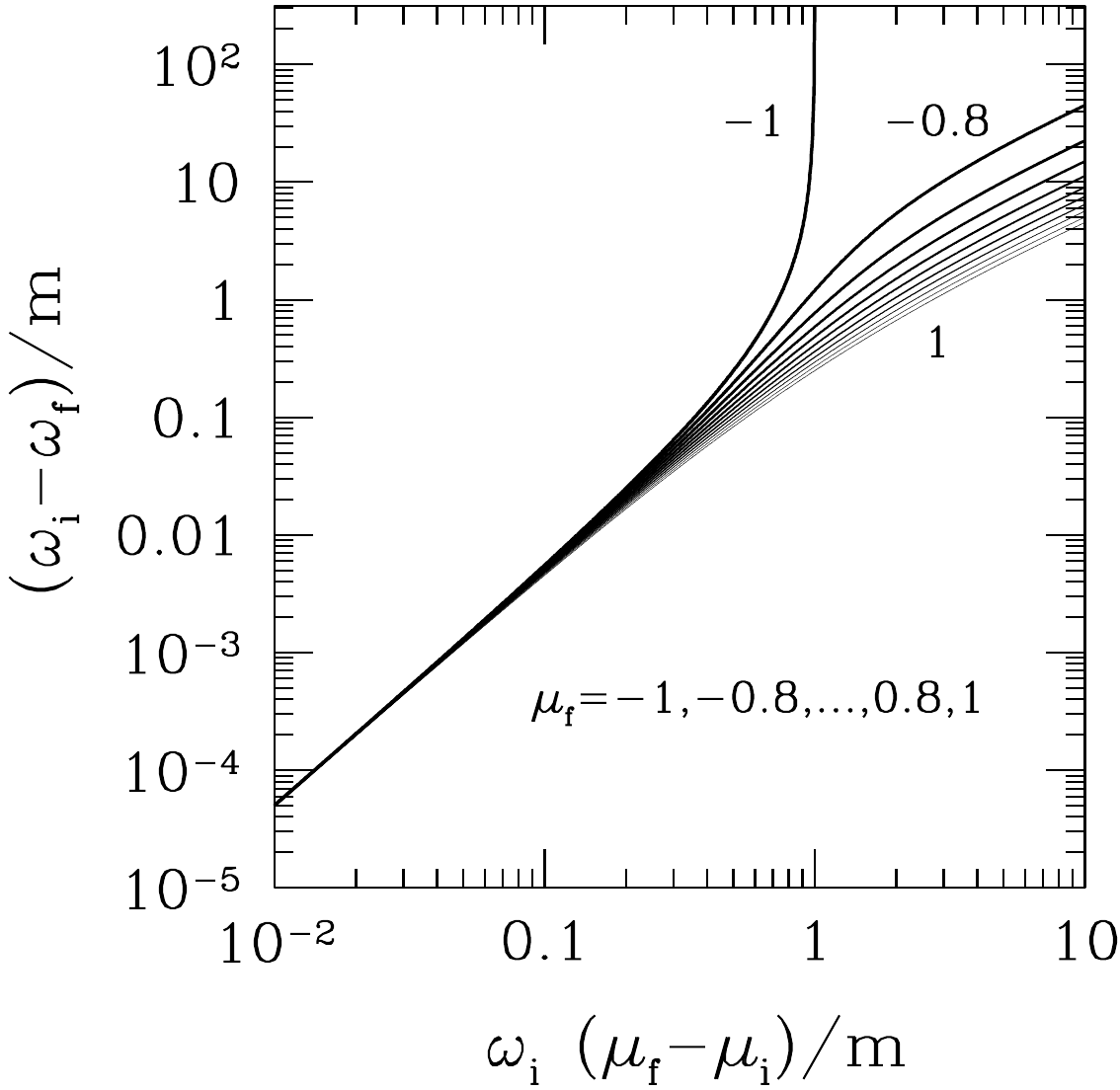}
\caption{Photon frequency shift by electron scattering.  Curves correspond to final direction cosine $\mu_f$
ranging from $-1$ to $1$ in steps of 0.2.}
\label{ComptonKinematicsFig}
\end{figure}

The kinematic relation between the initial and final electron states is modified compared with the vacuum case, 
because kinetic momentum is conserved only in the direction parallel to ${\bm B}$.  (As will be the case throughout
this paper, the labeling of initial and final particle states is contained in the accompanying Feynman diagrams.)  
The photon frequency shift following scattering off an electron at rest (from direction cosine $\mu_i = \cos\theta_i$ to
$\mu_f = \cos\theta_f$) is
\be\label{ComptonOmFin}
\omega_f - \omega_i = \frac{1}{1-\mu_f^2}\left ( \omega_i(\mu_f-\mu_i)\mu_f + m -\sqrt{\omega_i^2(\mu_f - \mu_i)^2 + 2m\omega_i(\mu_f - \mu_i)\mu_f + m^2} \right).  
\ee
This is derived by invoking the conservation of energy and longitudinal momentum, $\omega_i + m = 
\omega_f + (p_z^2 + m^2)^{1/2}$ and $\mu_i\omega_i  = \mu_f\omega_f + p_z$, to obtain the quadratic equation 
\be\label{eq:quadr}
(\omega_i + m - \omega_f)^2 = (\mu_i\omega_i - \mu_f\omega_f)^2 + m^2.
\ee
Although the outgoing photon frequency depends on three quantities ($\omega_i$, $\mu_i$ and $\mu_f$), the
frequency shift depends only on $\mu_f$ and $\omega_i(\mu_f-\mu_i)$ (Figure \ref{ComptonKinematicsFig}).
This expression reduces to $\omega_f \simeq \omega_i (1\pm \mu_i) / (1 \pm \mu_f)$  when $\omega_i|\mu_f-\mu_i| \gg m$,
with the upper (lower) sign corresponding to $\mu_f > \mu_i$ ($\mu_f < \mu_i$).

\subsection{Cross Section for $B \gg B_{\rm Q}$}

There is a considerable simplification in the electron-photon scattering cross section when Landau resonances can
be neglected in both intermediate and final electron/positron lines.  The formula for the cross section
becomes weakly dependent on background magnetic field strength $B$, for the simple reason that an electron
begins to behave like a ``bead on a wire.''  In the case where the initial electron is at rest, we find for 
the differential cross section
\be\label{eq:diffsig}
{1\over 2\pi}\frac{d\sigma}{d \mu_f} =  {r_e^2\over 2} 
\frac{\omega_f}{\omega_i} \frac{|F|^2 m^2}{E_f(E_f+m)}
{ |\varepsilon_i^z|^2 |\varepsilon_f^z|^2 \over 1-\beta_f \mu_f},
\ee
where $r_e = e^2/4\pi m = \alpha_{\rm em}/m$ is the classical electron radius, $\vec\varepsilon_{i,f}$ labels
the unit electric vector of the incoming and outgoing photons, $\beta_f = p_{z,f}/E_f$, and
\be\label{eq:ffactor}
|F| \equiv \left| \frac{4m(2m+\omega_i - \omega_f)}{[2m + \omega_i ( 1-\mu_i^2)]\,[2m - \omega_f ( 1-\mu_f^2)]} \right|.
\ee

\begin{figure} 
\epsscale{1.08}
%\plottwo{dsigma_mui_0PDF.pdf}{dsigma_mui_0p5PDF.pdf}
\plottwo{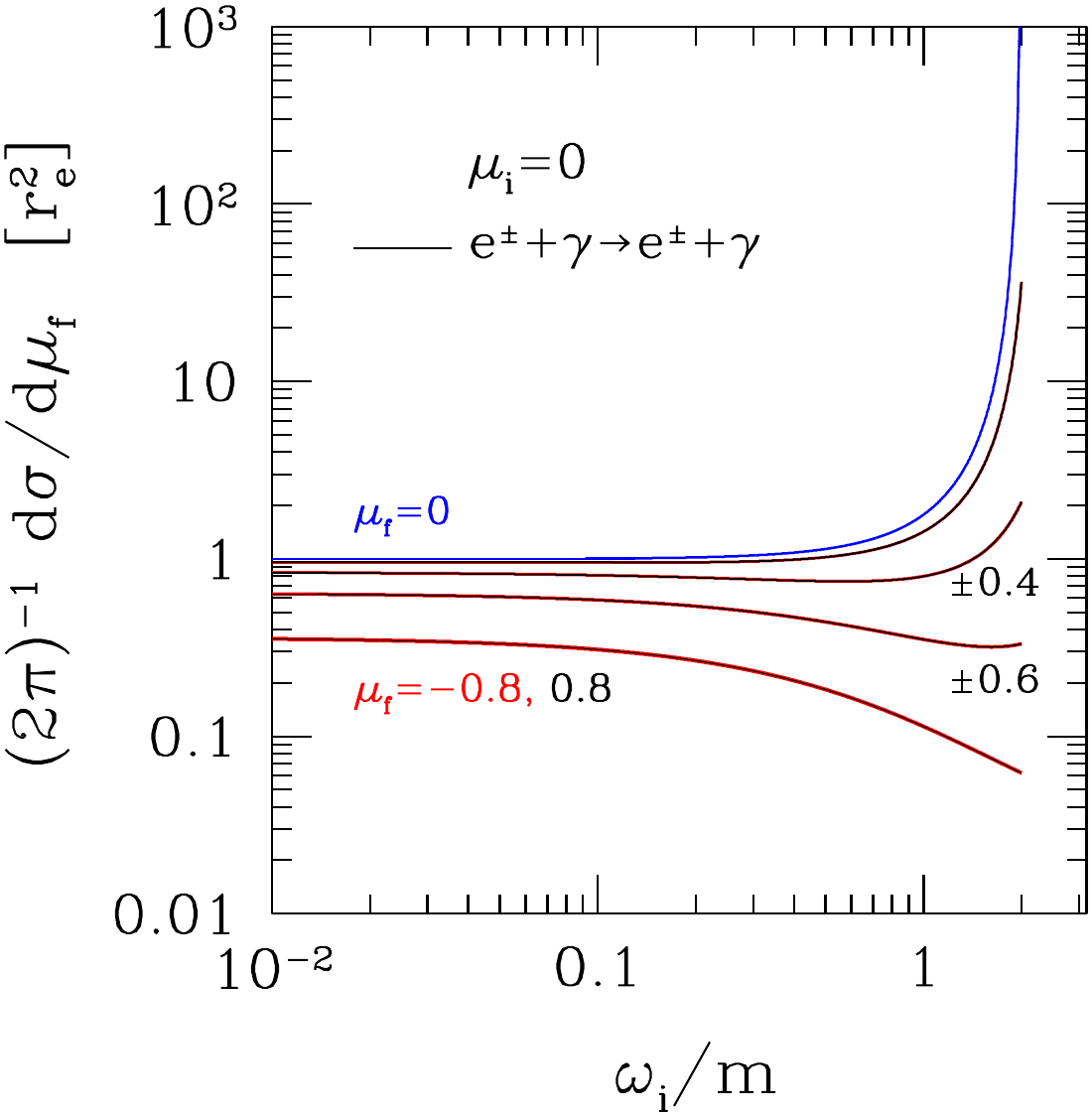}{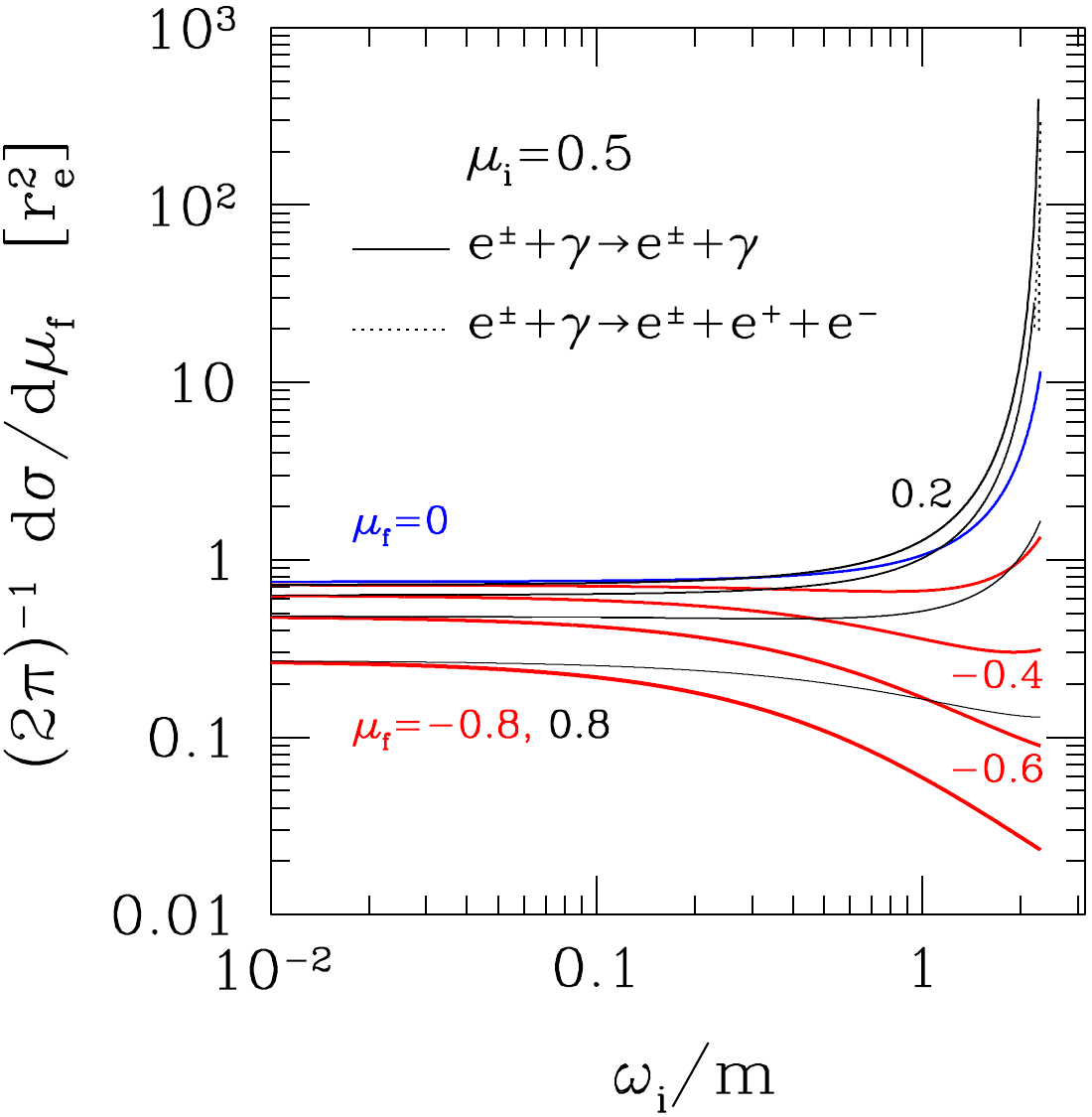}
%\includegraphics[width=\linewidth]{dsigma_mui_0PDF.pdf}
%\includegraphics[width=\linewidth]{dsigma_mui_0p5PDF.pdf}
%\plottwo{dsigma2_mui_0PDF.pdf}{dsigma2_mui_0p5PDF.pdf}
\plottwo{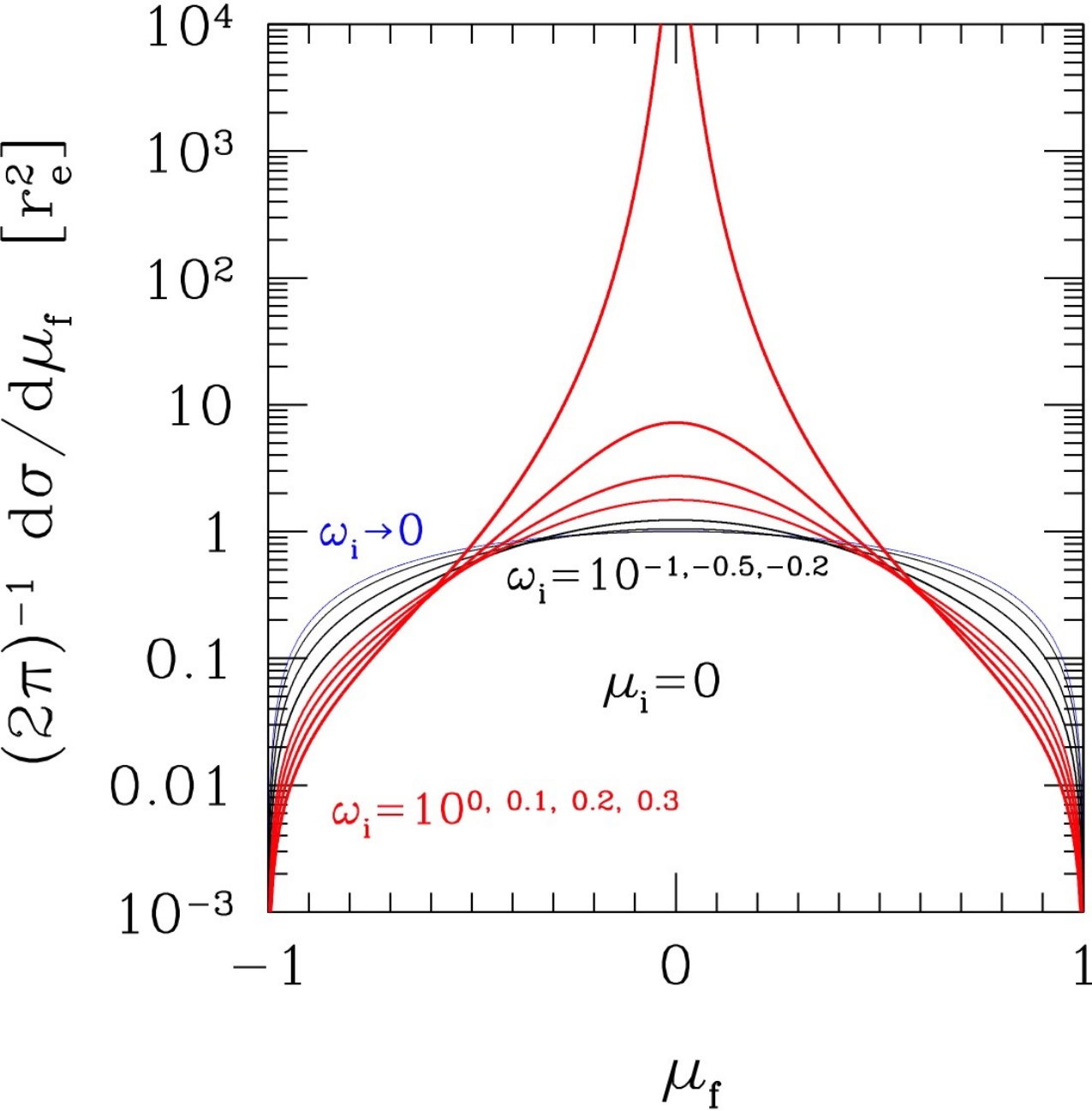}{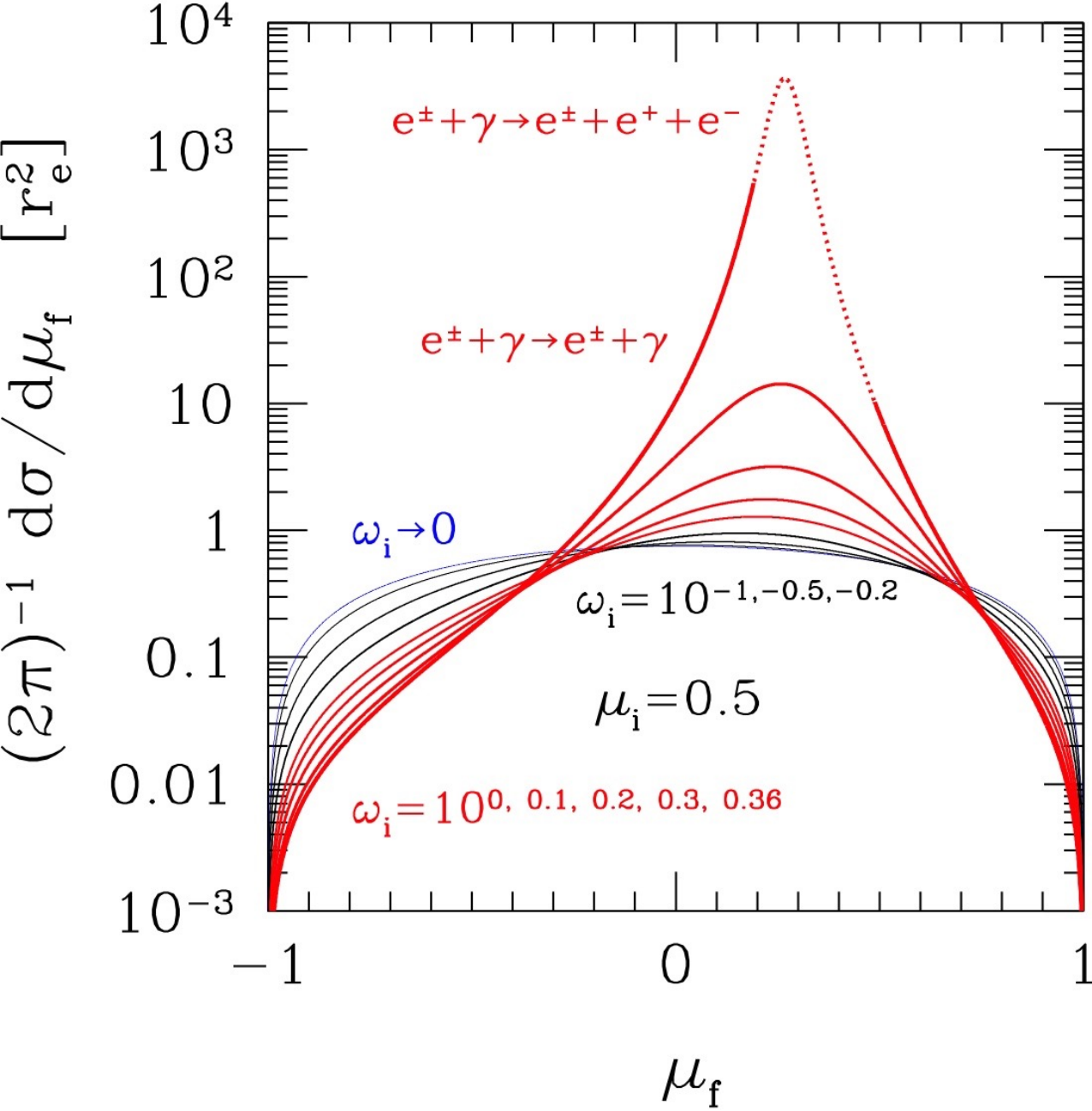}
\caption{Differential nonresonant electron scattering cross section versus initial photon frequency $\omega_i$
(top panels) and final photon direction cosine $\mu_f$ (bottom panels), in the regime $B \gg B_{\rm Q}$.  
Curves in the top panels represent
$|\mu_f| = 0, 0.2, 0.4, 0.6$.  Black curves: $\mu_f > 0$;  blue curve:  $\mu_f = 0$;  red curves:  $\mu_f < 0$.  
The cross section rises as the scattered photon approaches the threshold for single-photon pair creation.  
In the bottom panels, the blue curve marks the classical limit, and black and red curves are used for clarity.  Dotted curve:
the final-state photon rapidly converts to an electron-positron pair.}\label{NRDiffSig}
\end{figure}

The differential cross section is plotted in Figure \ref{NRDiffSig}.  The low-frequency behavior is consistent with
the classical result \citep{Canuto1971}
\be
{1\over 2\pi}\frac{d\sigma}{d\mu_f} = r_e^2 |\varepsilon_i^z|^2 |\varepsilon_f^z|^2 = r_e^2 \sin^2\theta_i\,\sin^2\theta_f.
\ee
The high-frequency behavior is more interesting.
Even when the initial photon energy lies well below the first Landau resonance, the scattering cross section 
spikes (but does not diverge) at a value of $\omega_f$ (and therefore $\omega_i$) somewhat larger than $2m$.  This 
spike arises from the pole in the matrix element associated with annihilation of the initial electron with a virtual
positron into the final-state photon \citep{Herold1979, DH1986}.  It appears at a lower photon frequency than
the first Landau resonance (energy $(m^2 + 2|e|B)^{1/2} - m$) if $B > 4B_{\rm Q}$.  

\begin{figure}
\epsscale{0.75}
%\plotone{sigma_scPDF.pdf}
\plotone{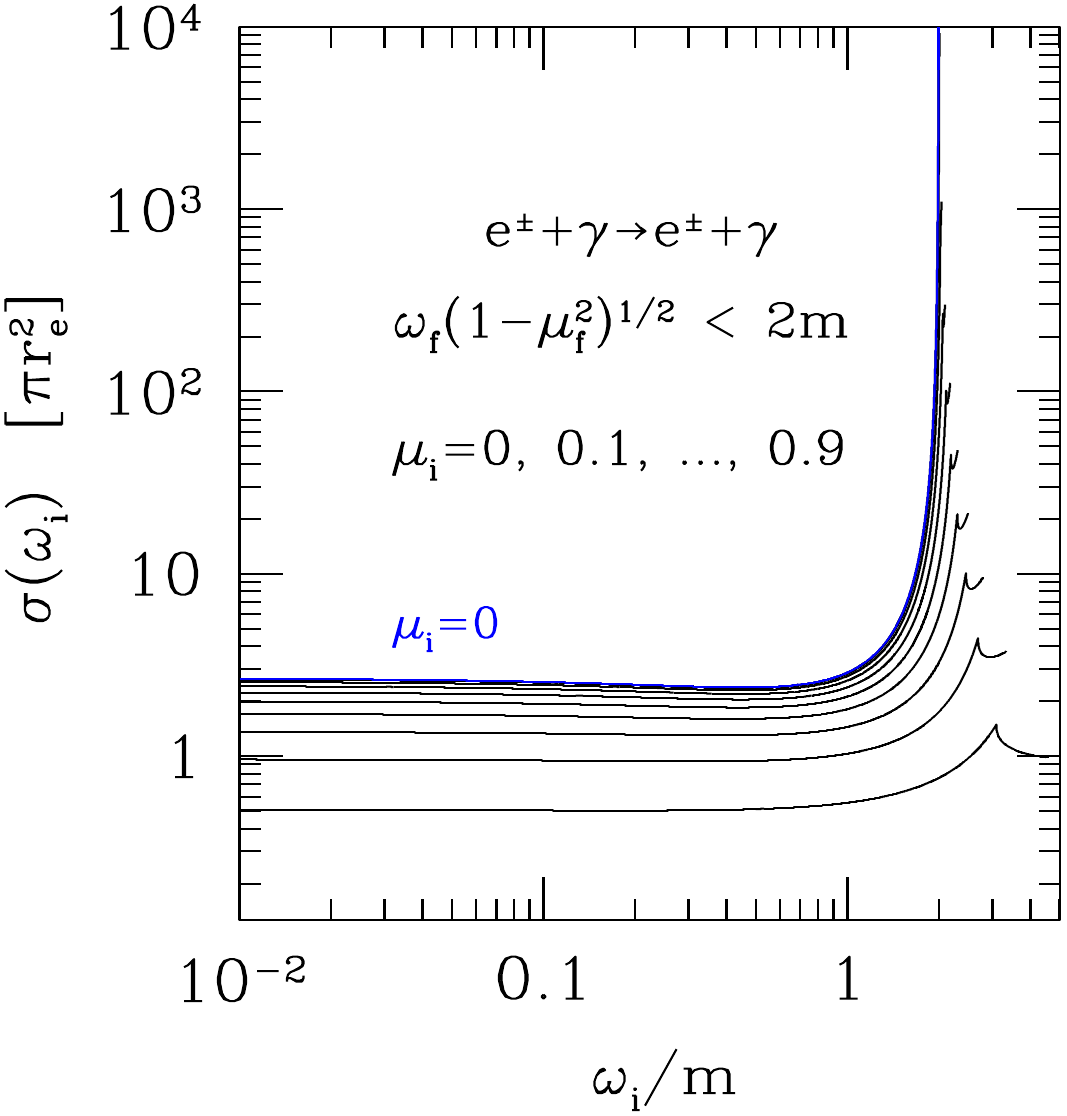}
\caption{Integral nonresonant electron scattering cross section plotted versus initial photon frequency $\omega_i$
for a range of initial direction cosine $\mu_i$, in the regime $B \gg B_{\rm Q}$.  The high-frequency peak marks the 
opening up of pair conversion in the final state (see Figure \ref{ComptonAssistedIntSig}).}
\label{NRIntSig}
\end{figure}

The partial cross section for scattering into non-pair-creating states is shown in Figure \ref{NRIntSig}.
The high-energy behavior of the total cross section is opposite to the Klein-Nishina result
for scattering in vacuum:  the cross section grows as  $\omega$ rises above $m$.   Nonetheless, scattering 
is still suppressed for photons that have a large energy owing to relativistic motion of the scattering charge along
${\bm B}$ (Lorentz factor $\gamma \gg 1$), because the $\sin^2\theta_i$ factor decreases as $\sim 1/\gamma^2$.

The high-frequency dip in the curves shown in Figure \ref{NRIntSig} represents the
opening up of the final-state phase space to pair creation over some range of scattering angles.   
A scattered photon can convert directly to an electron-positron pair if $\omega_f \sin\theta_f > 2m$, meaning that
the cross section for scattering-assisted pair creation, $e^\pm + \gamma \rightarrow e^\pm + e^+ + e^-$ can substantially
exceed the vacuum value.  This phenomenon is examined in more detail in Section \ref{PairCreation}.   

More generally, a high-frequency Klein-Nishina scaling for the scattering cross section -- which
is approached in the case of scattering at high-order Landau resonances in sub-QED magnetic fields \citep{Gonthier2000} --
loses meaning as $B$ rises above $4B_{\rm Q}$, because the scattered photon has a high probability of converting to a
pair \citep{BT2007}.

\vfil\eject
\subsection{Finite-{\rm B} Correction to the Cross Section}

\begin{figure} [h]
\epsscale{0.75}
%\plotone{sigma_sc_compNew.pdf}
\plotone{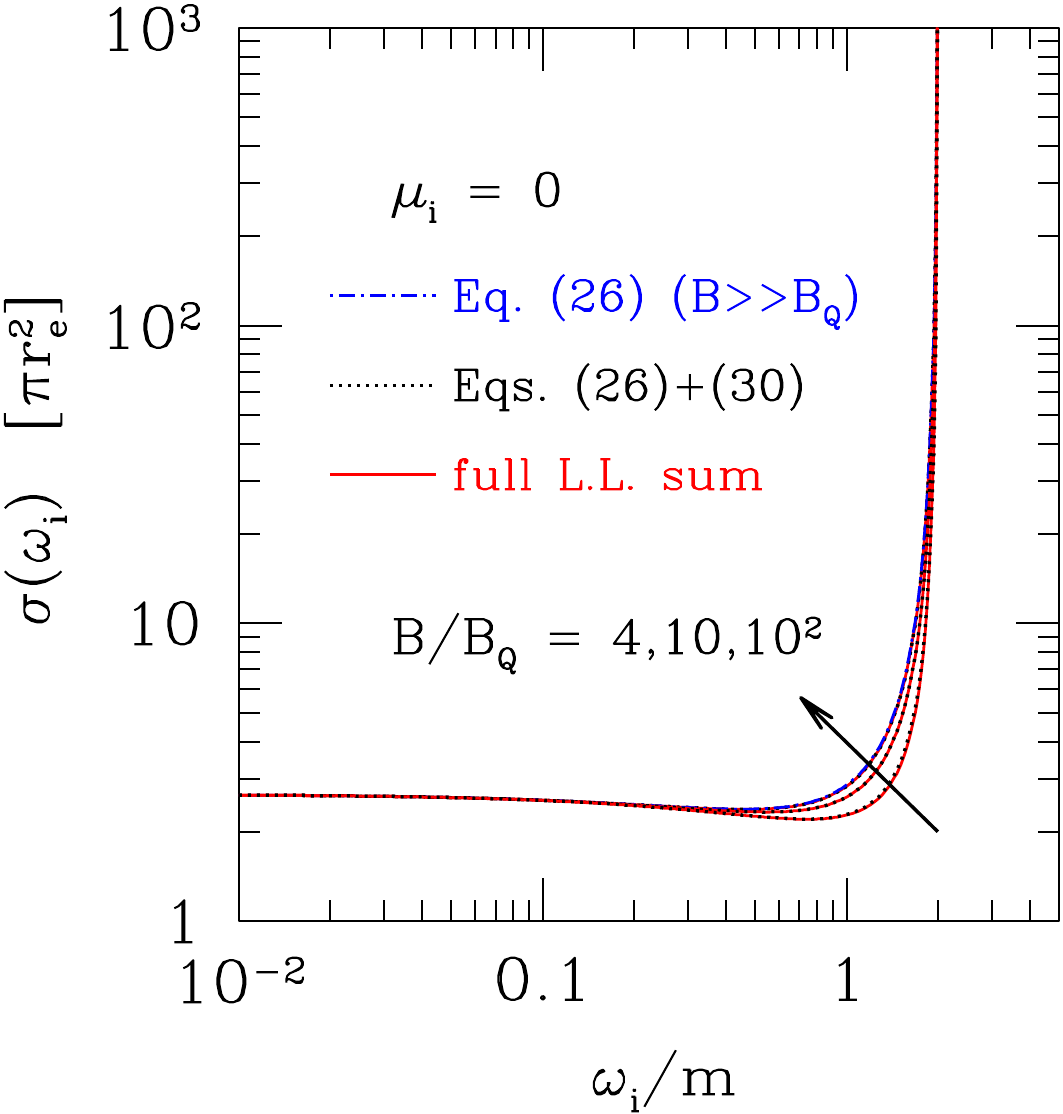}
\caption{Comparison of the exact cross section for electron-photon scattering from $n_i = 0$ to $n_f = 0$
(red curves) and the result obtained from truncated expansion in intermediate Landau level, with 
the additional correction factor (\ref{eq:corr}) included (overlying dotted black curves).  
Background magnetic field ranges from $B = 4B_{\rm Q}$ (the threshold value above which the first Landau
resonance is superseded by $e^\pm$ conversion) to $10B_{\rm Q}$ and $100B_{\rm Q}$.  Blue curve: 
the large-$B$ limit given by Equation (\ref{eq:diffsig}), which coincides very nearly with the uppermost curve 
($B = 100B_{\rm Q}$).}\label{fig:comp}
\end{figure}

A simple correction to the electron-photon scattering cross section representing a finite magnetic field is available.  
The overlap of the photon wave function (wavevector $k^\mu$) with a pair of harmonic oscillator wave functions,
such as appears in the scattering matrix, yields a factor $e^{-\lambda_B^2(k_x^2 + k_y^2)/4}$:
\ba\label{eq:phiint}
&& \int d^3x e^{i{\bm k}\cdot{\bm x}}\phi_n (x-a) \phi_0(x-b)e^{-iay/\lambda_B^2}
\left (e^{-iby/\lambda_B^2} \right)^* e^{ipz}\left ( e^{iqz} \right )^* \nn
&&\quad\quad = \frac{(2\pi)^2}{(2^n n!)^\frac{1}{2}} \delta\left ( k_y - \frac{a-b}{\lambda_B^2} \right ) 
\delta(k_z +p -q)e^{-\lambda_B^2 k_\perp^2/4}
e^{ik_x(b+\lambda_B^2 k_y/2)} \lambda_B^n (-k_y + ik_x)^n,\nn
\ea
as derived by \citet{DB1980}.  Including both photon vertices, the cross section is multiplied by
\be\label{eq:corr}
\sigma \rightarrow e^{-\lambda_B^2(k_{\perp,i}^2 + k_{\perp,f}^2)/2}\;\sigma; \quad\quad  k_\perp^2 = k_x^2 + k_y^2.
\ee
In Figure \ref{fig:comp} we compare the integral cross section derived from Equations (\ref{eq:diffsig}) and (\ref{eq:ffactor}),
with and without this correction, with the full sum over intermediate Landau states to be found in 
\citet{Herold1979}, \citet{MP1983B}, \citet{BAM1986}, \citet{DH1986}, and \citet{DH1991}.  There 
is very good agreement for $B = 100B_{\rm Q}$
in both cases, and excellent agreement for $B = 4B_{\rm Q}$ and $10B_{\rm Q}$ including the correction.

\subsection{Derivation}

Although the electron-photon scattering cross section in a strong magnetic field is already well covered in the literature,
we briefly review its derivation here and the Appendix.  Various components of the calculation will
find use in later sections, and some of the other processes considered (two-photon pair annihilation and creation)
are related by crossing symmetry.  The calculation is based on the rules summarized in Section \ref{s:rules}.

The cross section is obtained from the integral
\be\label{eq:sigscat}
\sigma = \frac{L^3}{T}\int L \frac{da_f}{2 \pi \lambda_B^2} \int L\frac{dp_{z,f}}{2 \pi}
\int L^3\frac{\omega_f^2d\omega_f d \Omega_f}{(2\pi)^3}\bigl|S_{fi}[1] + S_{fi}[2]\bigr|^2.
\ee
The initial electron is assumed to be at rest, so that the conservation of energy and longitudinal momentum 
are given by $m + \omega_i = E_f + \omega_f$ and $k_{z,i} = p_{z,f} + k_{z,f}$, leading to the recoil formula
(\ref{ComptonOmFin}).  

The two terms in the $S$-matrix correspond to the two diagrams in Figure \ref{NR}.  The first is
\be
S_{fi}[1] = -ie^2\int d^4x\int d^4x'\left [ \bar{\psi}_-^{(-1)}(x') \right ]_{p_{z,f},n_f,a_f} 
\gamma_{\nu} A^\nu(x)^* G_f(x'-x)\gamma_{\mu} A^\mu(x) \left [ \psi_-^{(-1)}(x) \right ]_{p_{z,i},n_i,a_i},
\ee
where $n_i = n_f = 0$, and the electron and photon wave functions are given in Secs. \ref{s:photon}-\ref{s:spinors}.
The second term in $S_{fi}$ is related to $S_{fi}[1]$ by an interchange of photon labels:  
$\bf{\omega_i \leftrightarrow -\omega_f}$ and $k_{z,i} \leftrightarrow -k_{z,f}$.  

Substituting for the electron propagator from Equation (\ref{ComptProp}), and restricting the sum over intermediate Landau
levels to $n_I = 0$, $\sigma = -1/+1$ (for electrons/positrons), the first term in $S_{fi}$ becomes
\be\label{eq:Sfi1}
S_{fi}[1] = \frac{-ie^2}{2\sqrt{\omega_i \omega_f} L^3}\left ( \frac{L}{2\pi} \right )^2  
2 \pi \delta (E_f + \omega_f -m -\omega_i)\int dp_{z,I} \int \frac{da_I}{\lambda_B^2}
\left( \frac{I_1 I_2}{m +\omega_i -E_I} + \frac{I_3 I_4}{m +\omega_i +E_I} \right).
\ee
Here,
\be\label{eq:I1}
I_1 = \int d^3x \left [ u_{0,a_I}^{(-1)*} ({\bm x})\right ]^T\gamma_0\gamma_{\mu}\varepsilon_i^{\mu}
e^{i{\bm k}_i\cdot{\bm x}}u_{0,a_i}^{(-1)}({\bm x})e^{i({\bm p}_i-{\bm p}_I)\cdot{\bm x}_\perp + i(p_{z,i}-p_{z,I})z}
\ee
and
\be\label{eq:I2}
I_2 = \int d^3x' \left [ u_{0,a_f}^{(-1)*}({\bm x}') \right ]^T\gamma_0\gamma_{\nu} \left (\varepsilon_f^{\nu}
e^{i{\bm k}_f\cdot{\bm x}'}  \right )^*u_{0,a_I}^{(-1)}({\bm x}')e^{i(-{\bm p}_f+{\bm p}_I)\cdot{\bm x}'_\perp + i(-p_{z,f}+p_{z,I})z'}.
\ee
The integral $I_3$ is obtained from $I_1$, and $I_4$ from $I_2$, by substituting the negative-energy wave function
$v^{(+1)}_{0,a_I}$ for the positive-energy wave function $u^{(-1)}_{0,a_I}$, and taking 
${\bm p}_I \rightarrow - {\bm p}_I$.

These integrals are evaluated the Appendix, where use is made of the overlap integral (\ref{eq:phiint}).
We find
\ba\label{eq:sfi1}
S_{fi}[1] &=& \frac{-ie^2}{2\sqrt{\omega_i \omega_f} L^5}\,
 e^{-ik_{x,f} (a_f - k_{y,f}\lambda_B^2/2)} \, e^{-\lambda_B^2(k_{i,y}^2+k_{\perp,f}^2)/4}\,
\frac{\varepsilon_i^z (\varepsilon_f^z)^*}{[2E_f(E_f+m)]^\frac{1}{2}} 
\frac{\omega_i (E_f + m) + p_{z,f} k_{z,i}}{(m+\omega_i)^2- E_I^2}\times \nn
&& \quad (2\pi)^3 \delta^{(3)}_{fi}(E,p_y,p_z)
\ea
where $E_I^2 = p_{z,I}^2 + m^2 = k_{z,i}^2 + m^2$.  Adding $S_{fi}[2]$ to this  and taking $\lambda_B \rightarrow 0$, one obtains a factor
\be\label{eq:feq}
F = \frac{\omega_i (E_f + m) + p_{z,f} k_{z,i}}{(m+\omega_i)^2 - m^2 - k_{z,i}^2} e^{-ik_{x,f}a_f} + 
\frac{-\omega_f (E_f + m) - p_{z,f} k_{z,f}}{(m-\omega_f)^2 - m^2 - k_{z,f}^2} e^{-ik_{x,f}a_i}.
\ee
Substituting $k_z = \mu \omega$, and making use of energy-momentum conservation,
we obtain the expression (\ref{eq:ffactor}) that appears in Equation (\ref{eq:diffsig}).  The differential cross section
is obtained by substituting $S_{fi}[1] + S_{fi}[2]$ into Equation (\ref{eq:sigscat}) and performing the $p_{z,f}$, $a_f$ and
$\omega_f$ integrals.  The factor $(1 - \beta_f \mu_f)^{-1}$ comes from integrating over the combination of the
$p_z$ and $E$ delta functions.

\section{Electron-Positron Pair Creation}\label{PairCreation}

A strong magnetic field opens up new efficient channels for converting gamma rays to electron-positron
pairs \citep{Erber1966, DH1983, Gonthier2000, BT2007}.  
The single-photon channel $\gamma \rightarrow e^+ + e^-$ is now consistent with
conservation of both momentum and energy, because the charges carry a generalized transverse momentum
$q{\bm A}$, and because the Lorentz invariance of the vacuum state is broken.  
The scattering of a sufficiently energetic gamma ray off an electron can also mediate pair creation through the
single-photon channel, achieving a high cross section over a narrow range of frequencies.  

Here, we are interested in the conversion of a photon (or photons) into a pair confined to the lowest Landau state.   The
conversion rate for a single photon is easily obtained using a detailed balance argument from the cross section for
single-photon annihilation, which was calculated by \citet{Wunner1979} and \citet{DB1980}.
We next consider the scattering of a photon below the Landau resonance into a pair-creating state, extending the
calculation of Section \ref{Compton}.  The cross section of this process
is strongly enhanced when the final-state photon resonates with the initial electron and a virtual positron (a $u$-channel 
resonance).  The cross section averaged over frequency significantly exceeds $\pi r_e^2$, and indeed can exceed the cross 
section for scattering at the first Landau resonance (where the final-state photon can directly convert to a pair only if 
$B > 4B_{\rm Q}$: \citealt{BT2007}).  

Importantly for the application to magnetars,
the cross section for photon collisions, $\gamma + \gamma \rightarrow e^+ + e^-$, is strongly enhanced compared
with the unmagnetized vacuum, by a factor of $\sim B/B_{\rm Q}$.  A calculation including the full intermediate Landau 
level sum can be found in \citet{KM1986}, and the regime $B \gg B_{\rm Q}$ is addressed by \citet{CT2008}.  
We consider the strong-field regime more fully here.  The kinematic threshold for photon collisions differs significantly
from the unmagnetized case, and the cross section is further enhanced when one of the colliding photons has a low frequency.

Our main focus is on the conversion of O-mode photons to pairs.  E-mode photons couple weakly to electrons by
scattering, with a cross section that is suppressed by a factor of $\sim (B/B_{\rm Q})^{-2}$ at energy $\sim m$.
The cross section for a collision with a second photon is similarly suppressed.   High-energy
E-mode photons are further depleted by splitting into two photons \citep{Adler1971}, a process that is kinematically allowed
for the E-mode but forbidden for the O-mode \footnote{Splitting of one photon into two daughters only conserves energy
and momentum if the index of refraction of one or both of the daughter photons is larger than the index of refraction of
the initial photon.   Since $n_{\rm O} > n_{\rm E}$ in magnetic fields both larger and smaller than $B_{\rm Q}$, only splitting of
the E-mode is allowed (e.g. \citealt{HL2006}).}.

\subsection {Single-photon Pair Creation into the Lowest Landau Level} \label{1gamma}

The energy of a photon cannot be reduced arbitrarily by a Lorentz boost in the presence of a background
magnetic field, because only a boost parallel to ${\bm B}$ leaves the background invariant.   The energy
is minimized in the frame where ${\bm k}\cdot {\bm B} = 0$;  hence, the threshold energy for pair creation is
\be\label{eq:eth}
\omega_{\rm min} = {2m\over \sin\theta}
\ee
for the ordinary polarization mode. 

To obtain the conversion rate, we start with the cross section of the inverse process of single-photon 
annihilation of an electron and positron \citep{Wunner1979, DB1980},
\be\label{eq:sig1gam}
{2 |p_z|\over E} \sigma_{\rm ann}  = 2\pi^2 {\alpha_{\rm em}\over E^2} {B_{\rm Q}\over B} e^{-2(B_{\rm Q}/B)^2(E/m)^2}.
\ee
This is evaluated in the center-of-momentum frame, with electron/positron momenta $\pm p_z$ and kinetic energies
$(p_z^2 + m^2)^{1/2}$, corresponding to perpendicular propagation of the photon.  The annihilation rate 
per electron is suppressed by a factor of $\sim B_{\rm Q}/B$ compared with vacuum (where $(2|p_z|/E)\sigma_{\rm ann} 
= \pi r_e^2$ in the nonrelativistic regime; \citealt{LLT4}), because the $e^\pm$ wave functions are concentrated
in an area of $\sim m^{-2} B_{\rm Q}/B$ transverse to the magnetic field.   

Consider now a thermal gas of pairs and photons at uniform temperature $T$ with densities
\be
{dn_\pm\over dp_z} = {|e|B\over (2\pi)^2} N_\pm;  
\quad\quad {d^2n_\gamma\over d\omega d\Omega} = {\omega^2\over (2\pi)^3} N_\gamma,
\ee
where $N_\gamma = (e^{\omega/T}-1)^{-1}$, $N_\pm = (e^{E/T}+1)^{-1}$.  Then detailed balance implies the
following relation for the decay rate $\Gamma_\pm$ into a pair:
\be\label{eq:db}
\Delta \omega \,\Delta\Omega {d^2n_\gamma\over d\omega d\Omega}\Gamma_\pm \cdot (1-N_\pm)^2 =
\left[\Delta p_z {dn_\pm\over dp_z}\right]^2 {2|p_z|\over E_p}\sigma_{\rm ann} \cdot (1+N_\gamma).
\ee
Substituting $\omega = 2E$ and the Jacobian factor $\Delta\omega\,\Delta\Omega/\Delta p_z^2 = 4\pi |p_z|/E\omega$, and 
boosting to a general frame gives
\be\label{eq:gampair}
\Gamma_\pm(\omega,\theta) 
= 2\alpha_{\rm em} {B\over B_{\rm Q}}  {m^4\over \omega_\perp^2(\omega_\perp^2-4m^2)^{1/2}}
 e^{-(B_{\rm Q}/2B)(\omega_\perp/m)^2} \sin\theta; \quad\quad \omega_\perp = \omega\sin\theta.
\ee

One observes that the annihilation rate per unit volume is enhanced by a net factor of $\sim B/B_{\rm Q}$
compared with a thermal pair plasma in the absence of the magnetic field.  The pair creation rate
per photon is therefore enhanced by the same factor, because the photon phase space does not depend
directly on $B$.  The same conclusion applies for two-photon pair creation (Section \ref{2gammareview}).

\subsection{Compton-assisted Pair Creation}\label{s:assist}

\begin{figure}
\epsscale{0.75}
%\plotone{sigma_pNew.pdf}
\plotone{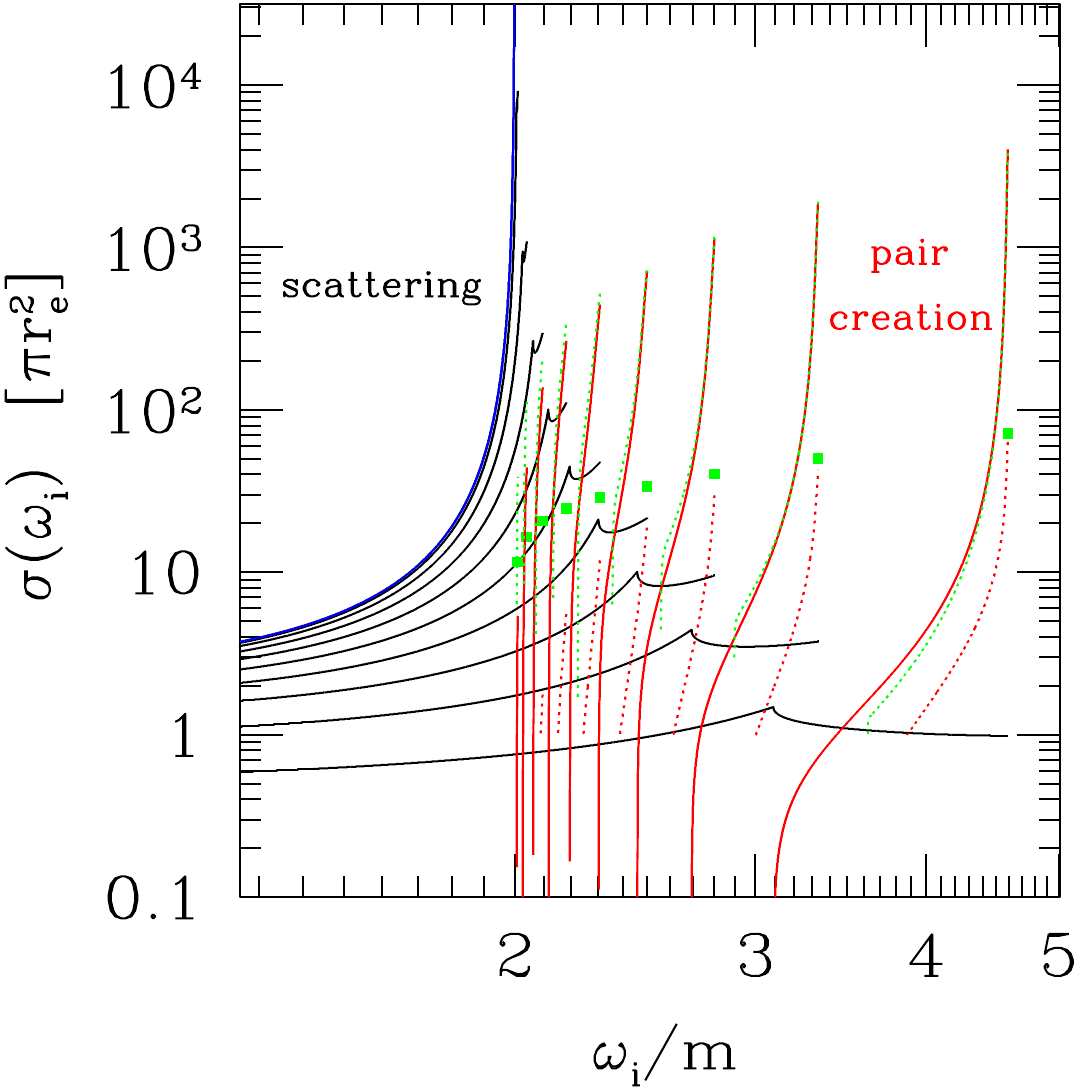}
\caption{Partial cross section for electron scattering into a non-pair-creating final state (black lines; see
also Figure \ref{NRIntSig}) and pair-creating final state (red solid lines).  Curves are ordered from left to
right by increasing $\mu_i = 0, 0.1, ... 0.9$.  Dotted red curves show the integral of the red curve over frequency
starting from low $\omega_i$.  Dotted green lines show the analytic approximation obtained by substituting
Equation (\ref{eq:poleapprox}) for the pole in Equation (\ref{eq:diffsig}), and green squares show the analytic approximation
(\ref{eq:sigint}) to the frequency-integrated cross section.}\label{ComptonAssistedIntSig}
\end{figure}

The scattering cross section (\ref{eq:diffsig}), as derived for a stationary target electron or positron, is
strongly peaked when the energy $\omega_f$ of the final-state photon approaches $2m/(1-\mu_f^2)$.  This exceeds
the threshold
(\ref{eq:eth}) for single-photon pair creation, meaning that electron scattering can effectively mediate pair creation,
$e^\pm + \gamma \rightarrow e^\pm + e^+ + e^-$.  Figures \ref{NRIntSig} and \ref{ComptonAssistedIntSig} show that the cross section peaks strongly for non-pair-converting final states when $\mu_i$ is small (the initial photon
propagates nearly perpendicular to the magnetic field) but peaks for pair-converting final states when
$\mu_i \rightarrow \pm 1$.  As we show here, the cross section can be well approximated analytically in the latter
regime.

\begin{figure} [h]
\epsscale{0.7}
%\plotone{pair_creationPDF.pdf}
\plotone{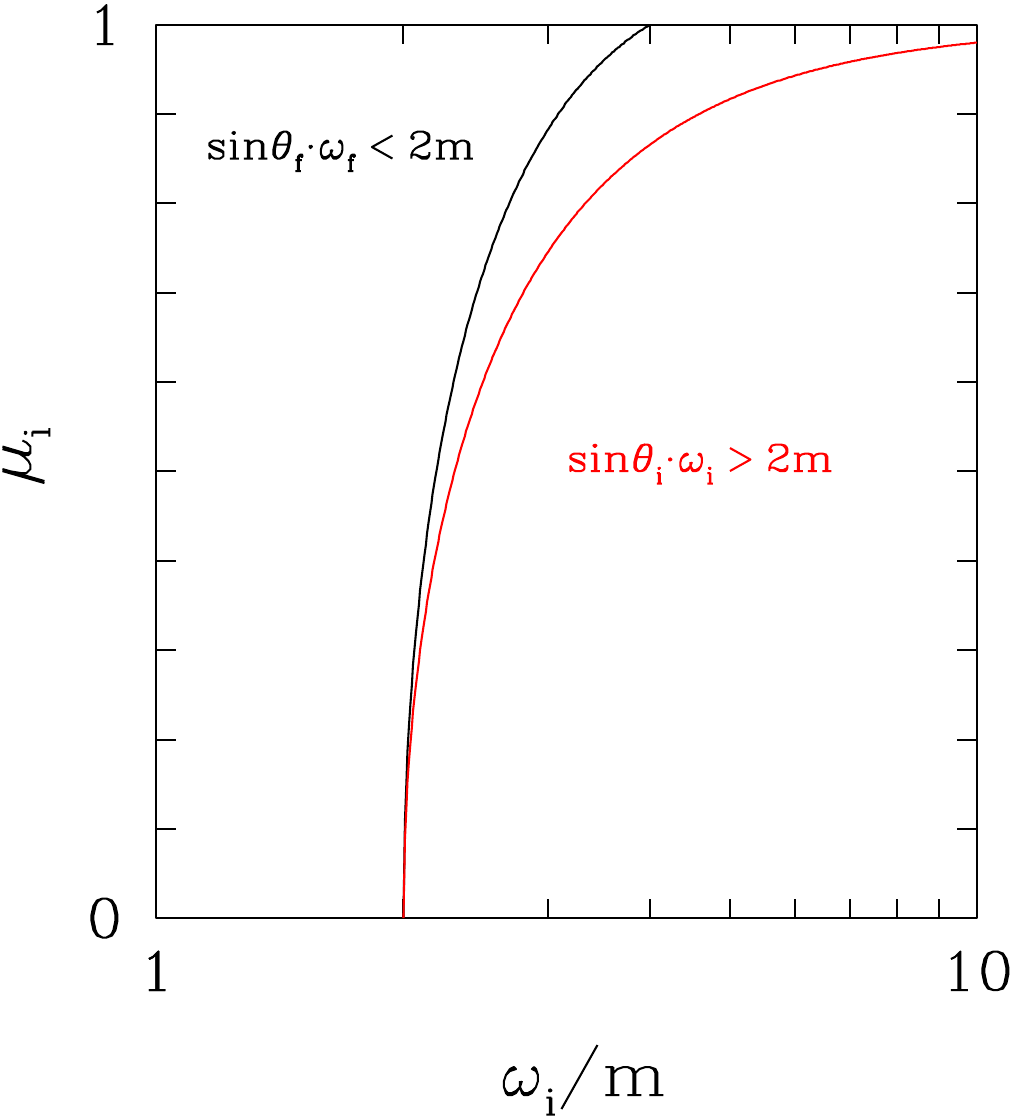}
\caption{A photon propagating in direction $\mu_i$ can be scattered into a pair-creating final state only
over a restricted range of initial frequency $\omega_i$.  The threshold $\omega_f(1-\mu_f^2)^{1/2} > 2m$ 
for single-photon pair creation is achievable in the final state only to the right of the black line 
(Equation (\ref{eq:comppair})); whereas the initial photon is itself below the threshold for pair creation only to the
left of the red line.  The range of frequencies where both conditions are satisfied grows wider as the propagation
direction becomes more aligned with ${\bm B}$.  In magnetar-strength magnetic fields, scattering-assisted pair
creation is possible at frequencies well below the threshold for resonant excitation to the first Landau
level.}\label{ComptonPCPhase}
\end{figure}

The initial energy $\omega_i$ of the photon must be large enough for it to have a chance at direct pair conversion
following scattering.  To obtain the minimum $\omega_i$ as a function of $\mu_i$, we first consider the final
direction cosine $\mu_f$ at which the perpendicular energy $s_f\omega_f$ is maximized.  
(Throughout this section, we use the shorthand $s_{i,f} \equiv \sin\theta_{i,f} = (1-\mu_{i,f}^2)^{1/2}$.)   
Differentiating Equation (\ref{eq:quadr}) with respect to $\mu_f$, one finds that this maximum occurs at 
$\mu_f = \mu_i \omega_i / (\omega_i+m)$ and is equal to 
\be
(s_f\omega_f)_{\rm max} = \sqrt{s_i^2\omega_i^2 + 2\omega_i m + m^2} - m.
\ee
Requiring $(s_f\omega_f)_{\rm max} > 2m$ gives the inequality $s_i^2\omega_i^2 + 2\omega_i m > 8m^2$,
which is satisfied for
\be\label{eq:comppair}
\omega_i > {\sqrt{1 + 8s_i^2}-1\over s_i^2} m.
\ee
This is shown as the black curve in Figure \ref{ComptonPCPhase};  in addition, the process of scattering-assisted
pair creation is interesting only if $s_i\omega_i < 2m$.  The threshold condition for pair creation is
$\omega_i > 4m$ in the case of nearly longitudinal propagation.

%The limiting values of $\mu_f$ for which conversion occurs are obtained by substituting 
%$(1-\mu_f^2)\omega_f^2 = 4m^2$ into Equation (\ref{eq:quadr}), giving
%\be\label{eq:mufth}
%\sin(\theta_f + \alpha) = \sin\beta;  \quad\quad \sin\alpha = {b\over \sqrt{a^2 + b^2}};  
%\quad\quad \sin\beta = {4m(\omega_i+m)\over \sqrt{a^2+b^2}}.
%\ee
%Here 
%\be
%a = 4m^2 + 2m \omega_i + s_i^2\omega_i^2; \quad\quad b = 4m\omega_i \mu_i.
%\ee
%(Throughout this section, we use the shorthand $s_{i,f} = \sin\theta_{i,f} = (1-\mu_{i,f}^2)^{1/2}$.)   
%Equation (\ref{eq:mufth}) has two solutions for $\mu_f$ if and only if the equation for $\beta$ itself has a solution,
%meaning that the initial photon must be more energetic than
%\be\label{eq:comppair}
%\omega_i > {\sqrt{1 + 8s_i^2}-1\over s_i^2} m.
%\ee

The pole in the cross section, seen in Equation (\ref{eq:ffactor}), is regulated by taking into account the width of the final-state photon, 
$\omega_f \rightarrow \omega_f - i\Gamma_\pm/2$, where the decay rate $\Gamma_\pm$ is given by Equation (\ref{eq:gampair}).
Then the denominator is replaced by
\be\label{eq:pole}
\left[(\omega_f - m)^2 - E_I^2\right]^2 \;\rightarrow\;  \left[(1-\mu_f^2)\omega_f^2 - 2m\omega_f\right]^2 + m^2\Gamma_\pm^2.
\ee
The first term on the right-hand side is minimized for a final-state photon energy and direction
\be\label{eq:finalpeak}
\omega_f^* = \omega_i + m - {m\over s_i};  \quad\quad \mu_f^* = \mu_i {s_i\omega_i - m\over s_i(\omega_i+m)-m},
\ee
corresponding to a final electron speed $\beta_f = p_{z,f}/E_f = \mu_i$.  The minimum value of this term is nonvanishing,
excepting when the initial photon approaches the threshold for one-photon pair creation, $s_i\omega_i \rightarrow 2m$, 
which also implies the pole condition $s_f^2 \omega_f \rightarrow 2m$.   Near the pole, we set 
$s_i\omega_i = 2m - \Delta\omega$, and the right-hand side of Equation (\ref{eq:pole}) can be approximated as
\be
 4m^2\left[\Delta\omega + m \left({1+s_i\over s_i}\right)^2\left(\mu_f - \mu_f^*\right)^2\right]^2 + m^2\Gamma_\pm^2.
\ee
Integrating over $\mu_f$, one finds that the pole in Equation (\ref{eq:diffsig}) yields the substitution
\ba\label{eq:poleapprox}
{1\over [2m - \omega_f(1-\mu_f^2)]^2} &\rightarrow&  {\pi s_i\over 2^{5/2}(1+s_i)}
  {\omega_f^2/m^{5/2} \over |\Delta\omega'|\,(|\Delta\omega| + |\Delta\omega'|)^{1/2}};\nn
|\Delta\omega'| &\equiv& \sqrt{\Delta\omega^2 + \Gamma_\pm^2/4},
\ea
with the other factors evaluated at final frequency and direction (\ref{eq:finalpeak}). With this replacement, we have an accurate formula for the integral cross section in the vicinity of the divergence;  see Figure \ref{ComptonAssistedIntSig}.

Fixing the initial photon propagation direction $\mu_i$, one can next average over $\omega_i$ by evaluating
the nonresonant factors in Equation (\ref{eq:diffsig}) at $\Delta\omega = 0$, to obtain the simple result
\be\label{eq:sigint}
\int d\omega_i \,\sigma(\omega_i,\mu_i) = 2\pi^2 r_e^2 m^{3/2} 
\left[\Gamma_\pm\left(\omega_f = {1+s_i\over s_i}m,\; \mu_f = {\mu_i\over 1+s_i}\right)\right]^{-1/2}.
\ee
This quantity appears as the green squares in Figure \ref{ComptonAssistedIntSig}.

To summarize, the cross section for scattering-assisted pair creation is greatly enhanced compared with the
vacuum value ($\sim \alpha_{\rm em} \cdot r_e^2$, see \citealt{LLT4}) owing to (i) a reduced energy loss to recoil 
by the scattered photon and (ii) the availability of rapid pair conversion following scattering.  (In other words, no 
additional vertex need be included in the diagram to obtain a pair in the final state.)

The frequency-averaged cross section peaks above Thomson, at a value given by Equation (\ref{eq:sigint}), which can be estimated as
\be
\bigl\langle\sigma\bigr\rangle(\mu_i) \sim {\int d\omega_i \sigma(\omega_i,\mu_i) \over 2m/s_i}
= \pi r_e^2 \cdot 10\left({B\over 10 B_{\rm Q}}\right)^{-1/2}.
\ee
This compares with the optical depth to resonant excitation at the first Landau level, which is 
\citep{Gonthier2000, BT2007}
\be
\bigl\langle \sigma\bigr\rangle = {1\over r}\int dr {2\pi^2 e^2\over mc} \delta\left(\omega - {|e|B\over mc}\right) 
\sim {2\pi^2 r_e^2\over 3 \alpha_{\rm em}(B/B_{\rm Q})} = \pi r_e^2 \cdot {30\over (B/10B_{\rm Q})}
\ee
in a dipolar magnetic field with $|d\ln B/d\ln r| \sim 3$.  In a magnetar-strength magnetic field, excitation to the
first Landau resonance generally requires relativistic bulk motion of the scattering charge along the magnetic field.

\subsection {Two-photon Pair Creation} \label{2gammareview}

\begin{figure} [h]
\epsscale{1.1}
%\plotone{2GammaPCLabelPDF.pdf}
\plotone{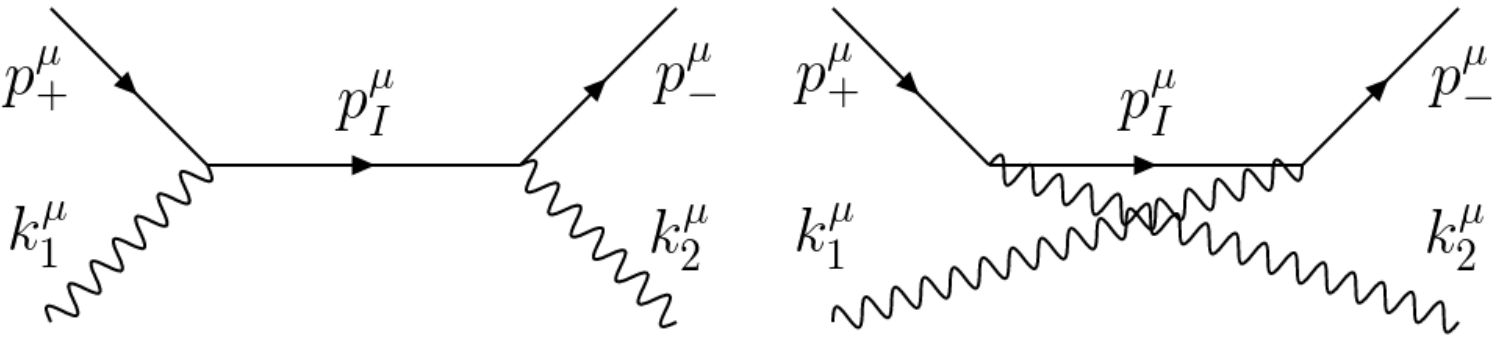}
\caption{Feynman diagrams for two-photon pair creation.}\label{2GPC}
\end{figure}

The collision of two photons to form an electron-positron pair (Figure \ref{2GPC}) occurs with a dramatically enhanced
cross section in an ultrastrong magnetic field.  One can see this using a detailed balance argument similar to the one
given in Section \ref{1gamma}.  The thermal equilibrium density of electrons and positrons is enhanced by a factor of
$\sim B/B_{\rm Q}$, whereas the density of photons is not.  Since the annihilation cross section is suppressed by 
one inverse power of $B/B_{\rm Q}$ \citep{DB1980}, the pair-production cross section grows by $\sim B/B_{\rm Q}$.

The cross section is evaluated to all orders in the intermediate-state Landau level by \citet{KM1986}, and the
special case of a longitudinal collision in an ultrastrong magnetic field is derived by \citet{CT2008}.  The
result does not depend on the choice of Dirac spinor basis as long as the spins of the electron and positron are
summed over.  Here, we consider the more general behavior of photon collisions in an ultrastrong magnetic field, 
where the pair is confined to the lowest Landau state, and outline the derivation using the spinor basis of Section
\ref{s:spinors}.

The kinematic constraints on photon collisions are altered by a magnetic field.  There is always a
Lorentz frame in which the total longitudinal momentum of the photons vanishes, $\mu_1 \omega_1 + \mu_2 \omega_2 = 0$.
In this frame, the threshold condition for pair creation is $\omega_1 + \omega_2 = 2m$.  One observes that
a photon of energy $\omega_1$ slightly less than $2m$ is able to collide with a much softer photon if the harder
photon moves approximately perpendicular to the magnetic field:  one requires only that $\omega_2 \geq |\mu_1|\omega_1$.
This contrasts with the unmagnetized vacuum, where $\omega_2 \geq m^2/\omega_1$, so that generally 
$\omega_1 + \omega_2 > 2m$.  Boosting along ${\bm B}$ to an arbitrary frame, the threshold condition is
\be
(\omega_1 + \omega_2)^2 - (\mu_1\omega_1 + \mu_2\omega_2)^2 > 4m^2 \quad\quad\quad (\gamma+\gamma \rightarrow e^+ +e^-).
\ee

The total cross section is, in the center-of-momentum frame where $p_{z,-} = -p_{z,+} = p_z$ and $E_+ = E_- = E$,
\be\label{eq:2gpair}
\sigma   = \frac{\pi}{\left | 1-\mu_{12} \right |} \left ( \frac{e^2}{4\pi m} \right )^2
\frac{B}{B_{\rm Q}}\frac{m^4}{\omega_1 \omega_2 E}
\frac{  |\varepsilon_1^z \varepsilon_2^z|^2}{|p_z|}   
\left |\frac{4p_z}{\omega_1\omega_2(1 - \mu_1 \mu_2 )^2 + 4\mu_1 \mu_2 p_z^2} \right |^2.
\ee
This lines up with the $B\gg B_Q$ limit of the result given by \citet{KM1986}.  

The cross section in an arbitrary Lorentz frame is obtained by expressing rest-frame quantities $\mu_i$ and $\omega_i$
in terms of the Lorentz scalars
$C_+ = (\omega_1 + \omega_2)^2 - (\mu_1 \omega_1 + \mu_2 \omega_2)^2$, $C_{ \perp i} = (1-\mu_i^2) \omega_i^2$ ($i = 1,2$),
and $C_\times = \omega_1 \omega_2 (1-\mu_1 \mu_2)$:
\ba\label{eq:2gpair2}
\sigma &=& {16\pi r_e^2\over k_1\cdot k_2} {B \over B_{\rm Q}} {\sqrt{C_+-4m^2}\over C_+^{1/2}}  
          {C_{\perp 1}C_{\perp 2}C_+^2 m^6\over
          [C_{\perp1}C_{\perp2}C_+ + 4m^2(C_\times^2 - C_{\perp1}C_{\perp2})]^2}\nn
       &=& {16\pi r_e^2\over |1-\mu_{12}|} {B\over B_{\rm Q}} {\sqrt{C_+-4m^2} \over C_+^{1/2} (\omega_1\omega_2)^3}   
          {C_+^2(1-\mu_1^2)(1-\mu_2^2) m^6\over [C_+(1-\mu_1^2)(1-\mu_2^2) + 4m^2(\mu_1-\mu_2)^2]^2}.
\ea
Here, $\mu_{12}$ is the relative direction cosine of the two photons, $k_1\cdot k_2 = \omega_1\omega_2(1-\mu_{12})$,
and $\mu_i$, $\omega_i$ are evaluated in an arbitrary frame.
The dominant correction to the cross section at finite $B$ comes from a factor
\be\label{eq:corr2}
\sigma \rightarrow e^{-\lambda_B^2(k_{\perp,1}^2 + k_{\perp,2}^2)/2}\;\sigma; \quad\quad  k_\perp^2 = k_x^2 + k_y^2.
\ee

Expression (\ref{eq:2gpair}) is accurate as long as the colliding photons do not propagate nearly parallel to 
${\bm B}$, e.g. as long as $(\varepsilon^z)^2 \gtrsim B_{\rm Q}/B$.   Otherwise, one has \citep{KM1986, CT2008}
\begin {equation}
\beta \cdot \sigma = {\pi r_e^2 \over 2 \omega^2}\frac{B}{B_Q}
\left | \frac{\varepsilon^+_2 \varepsilon^-_1 m^2}{(\omega+|p_{z}|)^2+m^2+ 2|e|B} -  
\frac{\varepsilon^+_1 \varepsilon^-_2 m^2}{(\omega-|p_z|)^2+m^2+ 2|e|B} \right |^2,
\end {equation}
once again as measured in the center-of-momentum frame where $\omega_1 = \omega_2 = \omega$ and $\beta = |p_z|/E$.

The Lorentz scalar $C_+$ measures how far the photons are above the threshold for pair creation.  For fixed $C_+$
the cross section exhibits a strong dependence on the energies of the colliding photons, as compared with the
Breit-Wheeler cross section as derived in an unmagnetized vacuum (see \citealt{LLT4}):
\begin {equation}
\sigma = \frac{\pi r_e^2}{2} (1-\beta^2)\left \{ (3-\beta^4) \ln\frac{1+\beta}{1-\beta} - 2 \beta (2-\beta^2) \right \}.
\end {equation}

\begin{figure}
\epsscale{0.75}
%\plotone{sigma2gPC1PDF.pdf}
\plotone{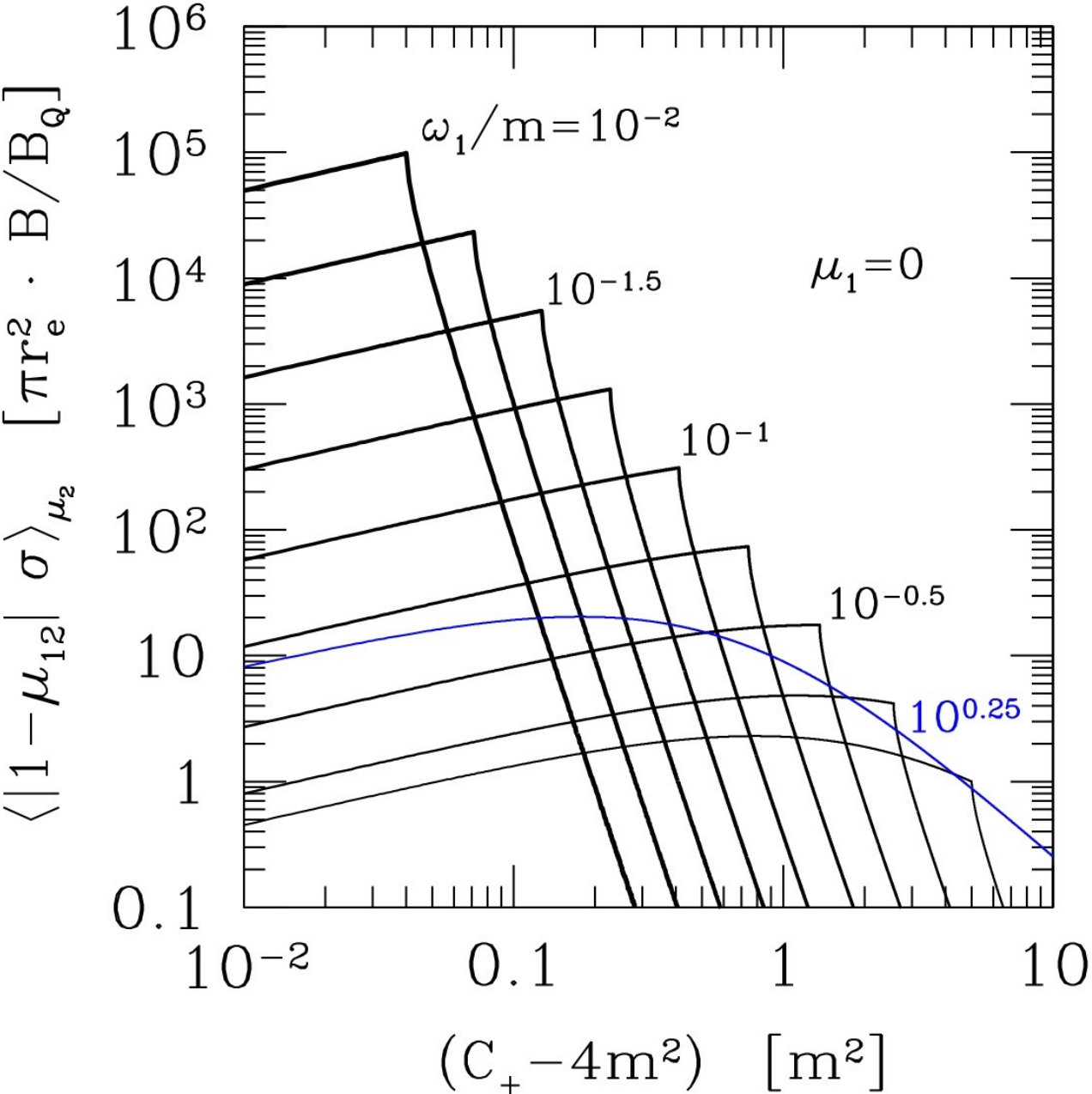}
\caption{Average of the photon collision rate over direction $\mu_2$ of the target photon, for a range
of photon energy $\omega_1$ and for $\mu_1 = 0$.}\label{2SigmaPC1}
\end{figure}

\begin{figure} [h]
\epsscale{0.75}
%\plotone{sigma2gPC2PDF.pdf}
\plotone{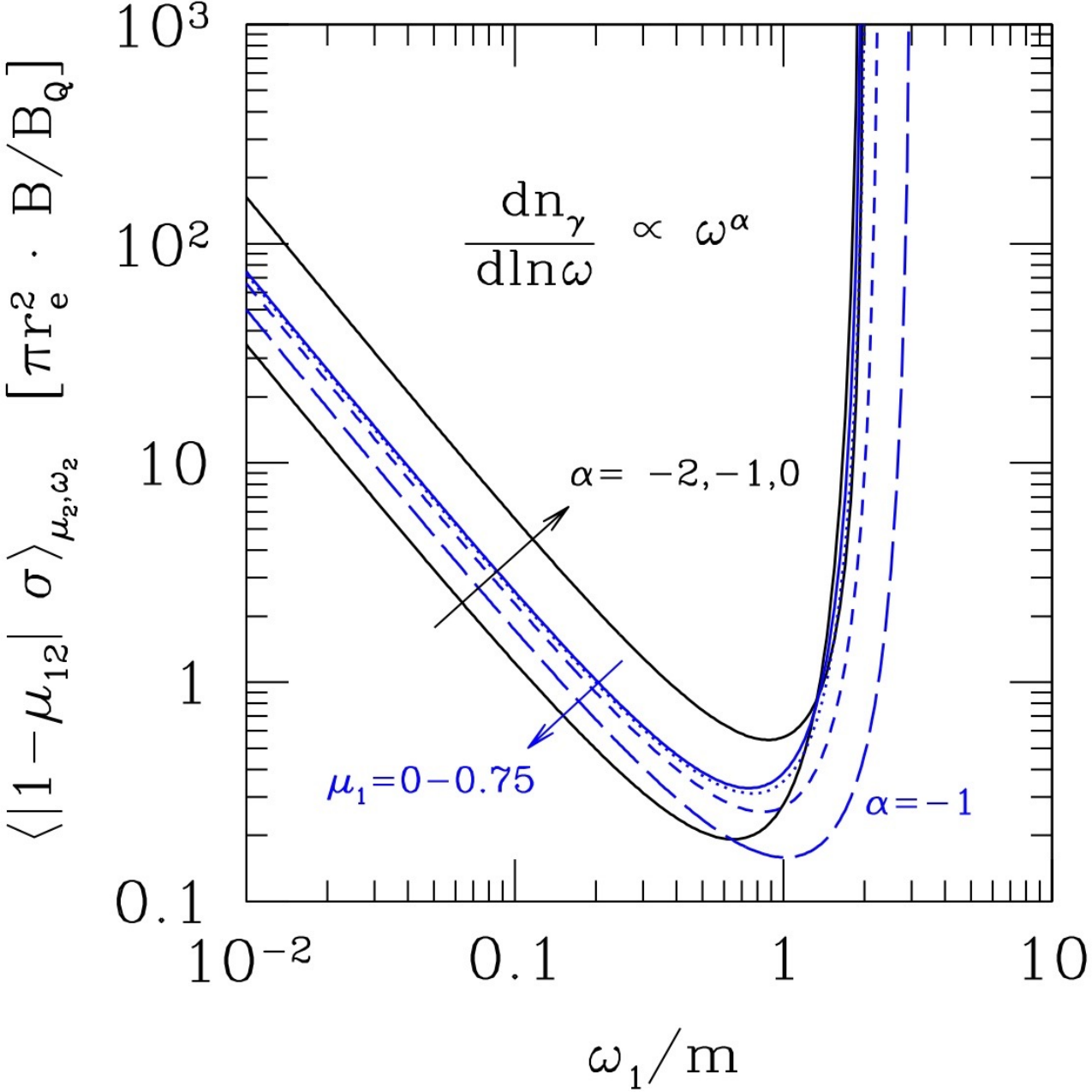}
\caption{Average of the photon collision rate over target photon direction and energy, weighted by the frequency
distribution of target photons, $dn/d\omega \propto \omega^{\alpha-1}$.  Black curves:  $\mu_1 = 0$ and 
$\alpha = -2$, $-1$, 0.  Blue dotted and dashed curves: range of $\mu_1$ and a flat energy spectrum 
($\alpha = -1$). Blue solid curve: $\alpha = -1$ and $\mu_1 = 0$. The rise at low frequencies and the second
rise near the threshold for single-photon pair creation both reflect the strong inverse frequency dependence of 
Equation (\ref{eq:2gpair2}), with this involving a low-frequency target photon in the latter case.}\label{2SigmaPC2}
\end{figure}

In particular, the cross section is significantly enhanced when one of the colliding photons has a low frequency
(Figure \ref{2SigmaPC1}).  The enhanced ability of a soft photon to remove a hard photon propagating perpendicular
to ${\bm B}$ implies that photons close to the threshold for single-photon pair creation are preferentially
removed by interactions with soft photons.  Averaging over a flat frequency
spectrum of the target photons (Figure \ref{2SigmaPC2}) also shows a strong divergence in
the averaged cross section as $\omega_1$ approaches the threshold for single-photon pair creation.

It is worth summarizing the different dependencies of the various pair creation channels on the strength of the background
magnetic field.  Whereas photon collisions grow more rapid as $B$ grows, the resonant and scattering-assisted channels
both get weaker.  In the last case, the weakening is not due to any change in the energy of the pole, but rather to
the increasing width of the scattered photon.  In magnetar-strength magnetic fields, the single-photon channel is
effectively instantaneous when the photon is above the kinematic threshold, and it will dominate two-body interactions.  
Which of the two-body effects most effectively removes hard photons that are below the single-photon conversion
threshold depends on the relative concentration of photons and pairs, as well as the strength of the magnetic field.

\subsubsection {Derivation}

The fastest route to a derivation of the cross section for $\gamma + \gamma \rightarrow e^+ + e^-$ is to make use
of crossing symmetry and infer the matrix element from that for electron scattering, $\gamma + e^\pm \rightarrow
\gamma + e^\pm$.  The cross section for pair production is then obtained from
\be\label{eq:sig2gam0}
\sigma = \frac{L^3/T}{\left | 1-\mu_{12} \right |}\int \frac{Lda_+}{2 \pi \lambda_B^2}\int \frac{Lda_-}{ 2 \pi\lambda_B^2 }
\int \frac{Ldp_{z,+}}{2 \pi}\int \frac{Ldp_{z,-}}{2 \pi}\sum_{\sigma_+,\sigma_-}\left| S_{fi} \right|^2.
\ee
Here, the outgoing electron and positron and the internal electron/positron lines are all restricted to the lowest 
Landau level ($\sigma_\pm = \pm 1$).
The matrix element for a process that has a particle $\phi$ with 4-momentum $p^\mu$ in the initial state is the same 
as the matrix element of a process that has that the antiparticle $\bar{\phi}$ with 4-momentum $-p^\mu$ in the 
final state \citep{PS1995}:
\be
S_{fi}(\phi(p^\mu) + ... \rightarrow ...) = S_{fi}( ... \rightarrow ... + \bar{\phi}(-p^\mu)).
\ee
Thus, the initial electron in the scattering process becomes a final positron, and the final photon a second 
initial photon.  Their momenta are related by
\be
p_i^\mu \rightarrow -p_+^\mu; \quad\quad k_f^\mu \rightarrow -k_2^\mu,
\ee
along with $p_f^\mu \rightarrow p_-^\mu$ and $k_i^\mu \rightarrow k_1^\mu$, following the labeling of states
shown in Figures \ref{NR} and \ref{2GPC}.

Before implementing this procedure, we must write down the scattering matrix element in a more general frame, where
the initial electron is not at rest.   One finds, after integrating over the intermediate-state delta functions in momentum,
the following generalization of Equation (\ref{eq:sfi1}),
\ba
S_{fi}[1] = {-ie^2\varepsilon^z_i\varepsilon_f^{z*}\over 2\sqrt{\omega_i\omega_f} L^5}
e^{-ik_{x,f}(a_f- \lambda_B^2k_{y,f}/2)} e^{-\lambda_B^2(k_{\perp,i}^2 + k_{\perp,f}^2)/4}
D(\omega_i,\omega_f,p_{z,i},p_{z,f}) \nn
\times (2\pi)^3 \delta^3_{fi}(E,p_y,p_z). 
\ea
Here,
\be
D = \frac{p_{z,i}[p_{z,f}(E_i+\omega_i+m) + p_{z,I}(E_f+m)] + (E_i+m)[(E_f + m)(E_i + \omega_i - m) + p_{z,f}p_{z,I}]}
           {2\sqrt{E_i E_f(E_i + m)(E_f + m)}[(E_i + \omega_i)^2 - E_I^2]},
\ee
and $p_{z,I} = p_{z,i} + k_{z,i}$, $E_I = \sqrt{p_{z,I}^2 + m^2}$.   The expression reduces to Equation (\ref{eq:sfi1})
after taking $p_{z,i} \rightarrow 0$.  Now applying the crossing symmetry and moving into
the center-of-momentum frame of the resulting pair, one finds for the pair-production matrix element
\ba
S_{fi}[1] = {-ie^2\varepsilon^z_1\varepsilon_2^z\over 2\sqrt{\omega_1\omega_2} E L^5}
e^{ik_{x,2}(a_2 + \lambda_B^2k_{y,2}/2)} e^{-\lambda_B^2(k_{\perp,1}^2 + k_{\perp,2}^2)/4}
{m(\omega_1\mu_1 - 2p_{z,+})\over 2\omega_1(\mu_1p_{z,+}-E) + \omega_1^2(1-\mu_1^2)} \nn
\times (2\pi)^3 \delta^3_{fi}(E,p_y,p_z).
\ea
(The freedom of choice of background magnetic gauge allows us to set $k_x = 0$ for one but not both of the
colliding photons.) Note also that this matrix element lines up with the $B\gg B_Q$ limit of the two-photon 
annihilation matrix element from \citet{DB1980}, as expected. The second exchange term in the matrix element 
(Figure \ref{2GPC}) is obtained by interchanging
$\varepsilon_1$, $k^\mu_1$ with $\varepsilon_2$, $k^\mu_2$.  Substituting $S_{fi}[1]+S_{fi}[2]$ into Equation 
(\ref{eq:sig2gam0}) and setting $B \gg B_{\rm Q}$ gives Equation (\ref{eq:2gpair}).

\section{Two-photon Pair Annihilation}\label{s:ann2g}

The cross section for the annihilation of an electron and positron into two photons is suppressed by a factor of $\sim 
(B/B_{\rm Q})^{-1}$ in the presence of an ultrastrong magnetic field.  The full sum over intermediate-state Landau levels
is presented by \citet{DB1980}.   The net rate of pair annihilation is further suppressed, as we quantify here, by
the reconversion of one or both of the created photons back to a pair. When a photon exceeds the energy threshold
(\ref{eq:eth}), this conversion to a pair is generally very rapid compared with any further two-particle interactions.
The net effect is to reduce the overall annihilation rate within a gas of electrons and positrons;  the strength of
the effect grows rapidly as the colliding pairs become mildly relativistic.   

We first present an integral formula relating the annihilation cross section to the cross section for two-photon 
pair creation, which takes a very simple form when  $B \gg B_{\rm Q}$.  
The annihilation cross section is given by the phase-space integral
\be\label{eq:sigann0}
\sigma_{\rm ann} = \left | \frac{p_{z,+}}{E_+} - \frac{p_{z,-}}{E_-} \right |^{-1}\frac{1}{2}\int \frac{L^3 \omega_1^2 d\omega_1d\Omega_1}{(2\pi)^3}\int \frac{L^3 \omega_2^2 d\omega_2d\Omega_2}{(2\pi)^3}
\int\frac{da_+}{L}\int\frac{da_-}{L}\frac{L^3}{T}\sum_{\sigma_+,\sigma_-} |S_{fi}|^2.
\ee
The incoming pair and intermediate $e^\pm$ lines are restricted to the lowest Landau state, $\sigma_\pm = \pm 1$.  
Two-photon pair annihilation and creation involve essentially the same matrix element, with all momenta simply reversed
in sign:
\be\label{eq:sfi_ac}
|S_{fi}|^2_{\rm ann} = |S_{fi}|^2_{\rm cre} = S_0^2(p_{z,-},p_{z,+},\omega_1,\omega_2,\mu_1,\mu_2)\cdot 
(2\pi)^3 \delta^3_{fi}(E,p_y,p_z).
\ee
We work in the center-of-momentum frame ($p_{z,-} = -p_{z,+} = p_z$, $E_+ = E_- = E$).
Substituting this into the phase-space integral (\ref{eq:sig2gam0}) gives a relation between $S_0$ and the
two-photon pair creation cross section (\ref{eq:2gpair}),
\be\label{eq:s0sq}
S_0^2 = {2\pi T\lambda_B^2 |p_z|\over L^8 E}\cdot |1-\mu_{12}| \sigma_{\rm cre}.
\ee
Combining Equations (\ref{eq:sfi_ac}) and (\ref{eq:s0sq}) with the integral (\ref{eq:sigann0}) gives for the annihilation
cross section
\be \label{eq:sigrelation}
\sigma_{\rm ann} = {\lambda_B^4\over 2}\int d\mu_1 d\mu_2 \left[ {\omega_1^2\omega_2^2\over |\mu_1-\mu_2|} \cdot
|1-\mu_{12}|\sigma_{\rm cre}\right].
\ee
A more explicit form for this cross section is obtained by substituting Equation (\ref{eq:2gpair}) for $\sigma_{\rm cre}$,
\be\label{eq:sigann}
\sigma_{\rm ann} = {2\pi r_e^2\over B/B_{\rm Q}} {\beta\over\gamma^2} \int d\mu_1 d\mu_2  
{-|\mu_1-\mu_2| (1-\mu_1^2)(1-\mu_2^2)\over \mu_1\mu_2 [(1-\mu_1\mu_2)^2 - \beta^2(\mu_1-\mu_2)^2]^2}.
\ee
Here, $\beta = |p_z|/E$ is the speed of the outgoing electron and positron, and
\be\label{eq:om12}
\omega_1 = \frac{2E \mu_2}{\mu_2 - \mu_1} ; \quad\quad \omega_2 = \frac{2E \mu_1}{\mu_1 - \mu_2}.
\ee

\begin{figure}
\epsscale{1.1}
%\plottwo{2GamAnnihilGam1p3VVVLargePDF.pdf}{2GamAnnihilGam2p5VVVLargePDF.pdf}
\plottwo{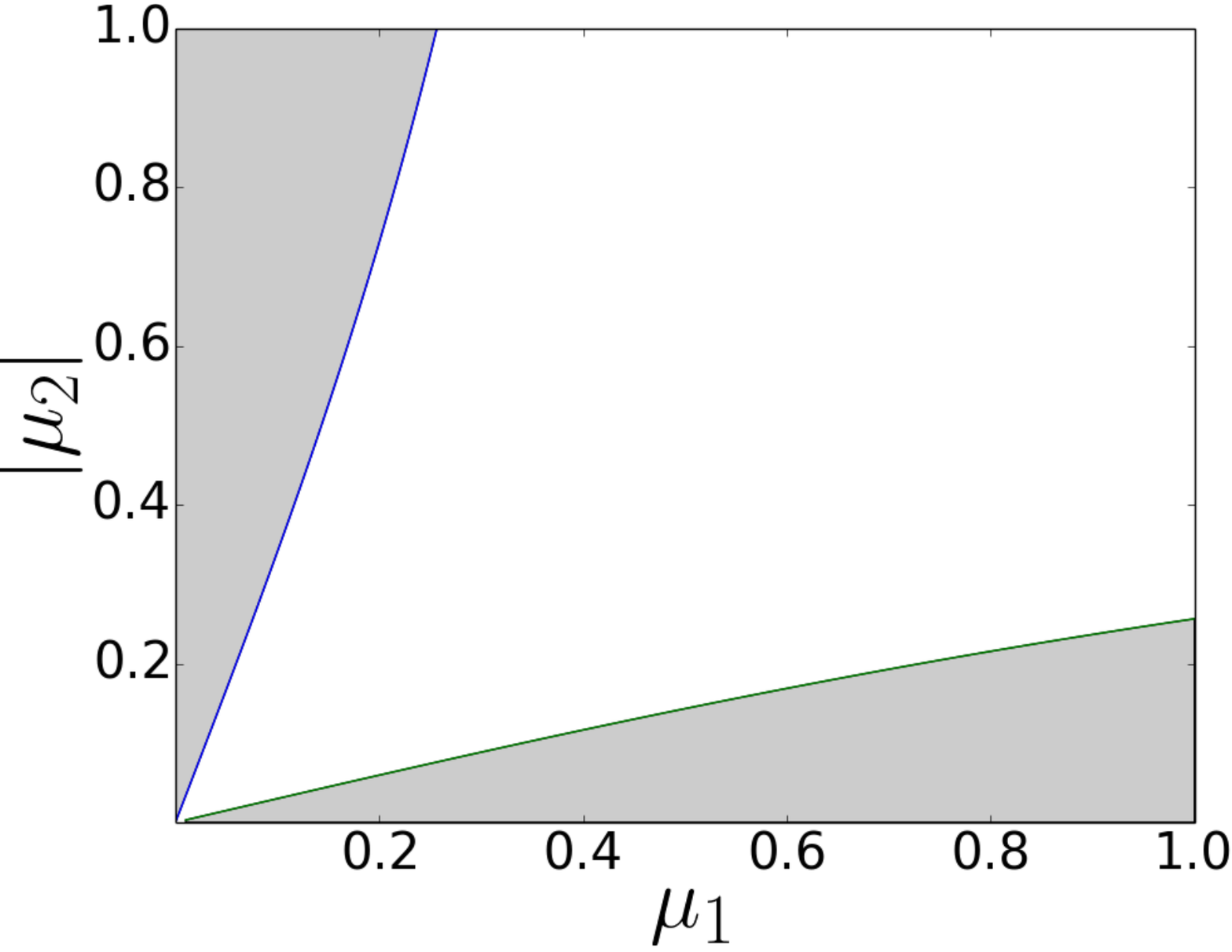}{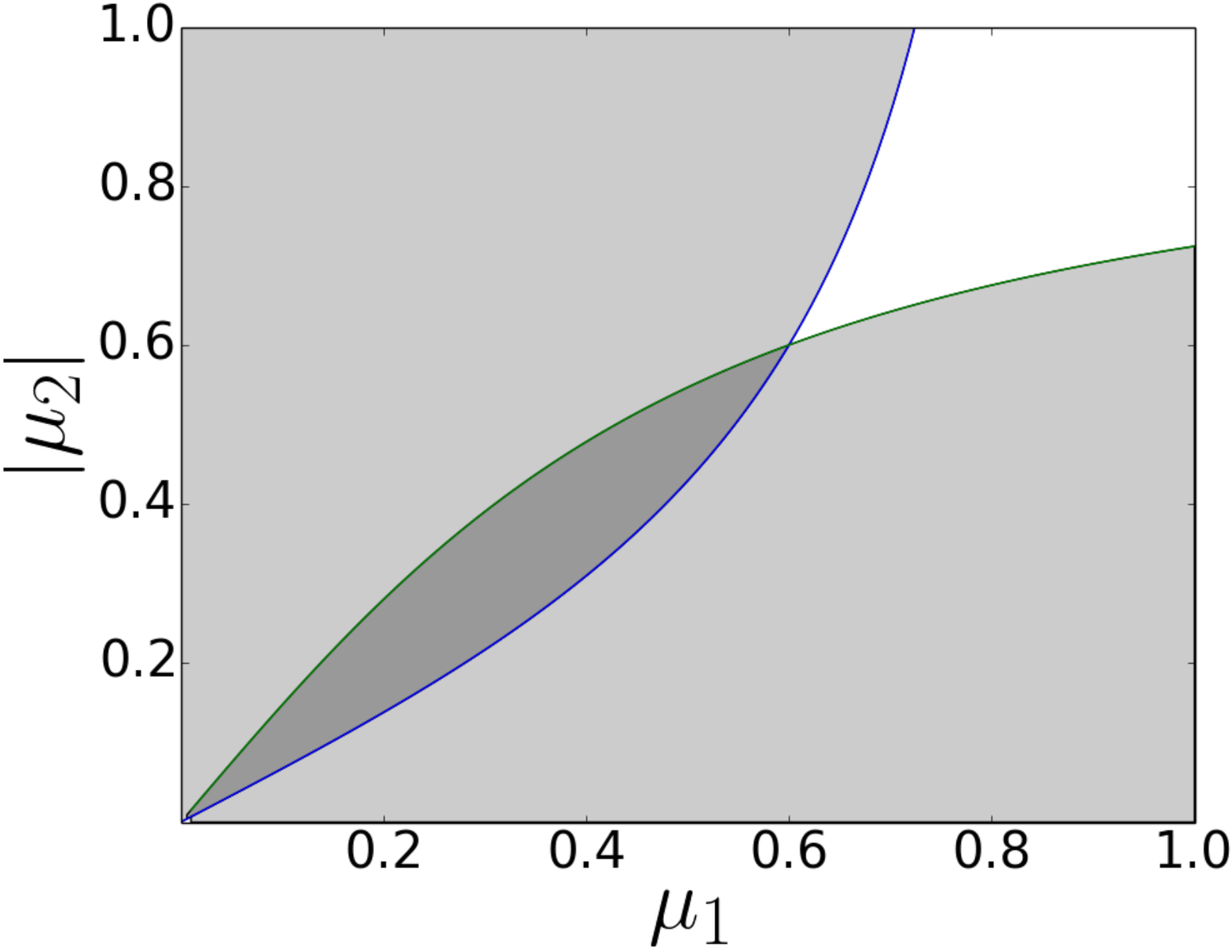}
\caption{In each panel, the left shading marks the zone where photon 1 ($\mu_1 \geq 0$) is above threshold for
pair conversion.  The right shading marks the zone where photon 2 ($\mu_2 \leq 0$) is above threshold.  These 
zones are excluded from the integral (\ref{eq:sigann}).
Left panel: colliding $e^\pm$ each have energy $\gamma = 1.3$ in the center-of-momentum frame.  Right panel:  
$\gamma = 2.5 > 2$;  both photons convert to a pair in the overlapping shaded zone.}  \label{2GamPhaseSpace}
\end{figure}

Demanding that both of the created photons remain below the threshold for pair creation,
$(1-\mu_i^2)^{1/2}\omega_i < 2m$, restricts the range of $\{\mu_1,\mu_2\}$ in the integral (\ref{eq:sigann}).
Since $\mu_1\cdot\mu_2 \leq 0$ is kinematically required
in the center-of-momentum frame, we focus on the quadrant $\mu_1 \geq 0$ and $\mu_2 \leq 0$.
Given a value of $\mu_1$, consider the threshold
for photon 2 to pair create, $\omega_2 (1-\mu_2^2)^{1/2} > 2m$.  Substituting Equation (\ref{eq:om12}) for $\omega_2$,
one obtains the range of $\mu_2$ in which two photons survive in the final state,
\be\label{eq:mu2th}
|\mu_2| >  \mu_{\rm th}(\mu_1) \equiv
{\mu_1(\gamma\sqrt{1+\mu_1^2\beta^2\gamma^2} -1)\over 1 + \gamma^2\mu_1^2} \quad\quad (0 \leq \mu_1 \leq 1).
\ee
There is a similar bound $\mu_1 > \mu_{\rm th}(|\mu_2|)$ to avoid pair creation of photon 1.   The excluded zones
in the space $\{\mu_1,|\mu_2|\}$ are marked out in Figure \ref{2GamPhaseSpace}.   

The possibility that both photons convert to pairs opens up when the line $\mu_1 = |\mu_2|$ first intersects the curve
(\ref{eq:mu2th}), which happens when $\gamma = 2$.  Then the entire zone
\be\label{eq:twopairs}
\mu_1, |\mu_2| < \mu_{\rm min} = \left(1-{4\over \gamma^2}\right)^{1/2}\quad\quad (\gamma > 2)
\ee
is excluded.  Combining these bounds and setting $\mu_{\rm min} = 0$ for $\gamma < 2$ gives the full integral
for final states in which both photons remain below threshold for conversion to a pair,
\be\label{eq:sigann2}
\sigma_{\rm ann} = {8\pi r_e^2\over B/B_{\rm Q}} {\beta\over\gamma^2}
\int_{\mu_{\rm min}}^1 {d\mu_1 \over \mu_1}
\int_{\mu_{\rm th}(\mu_1)}^{\mu_1} {d|\mu_2|\over|\mu_2|}
{(\mu_1+|\mu_2|) (1-\mu_1^2)(1-\mu_2^2)\over \bigl[(1+\mu_1|\mu_2|)^2 - \beta^2(\mu_1+|\mu_2|)^2\bigr]^2}.
\ee
Just as in the case of two-photon pair creation, the dominant finite-$B$ correction to this formula
comes from the exponential factor (\ref{eq:corr2}).

\begin{figure} [h]
\epsscale{0.75}
%\plotone{SigmaAnnihilCTPDF.pdf}
\plotone{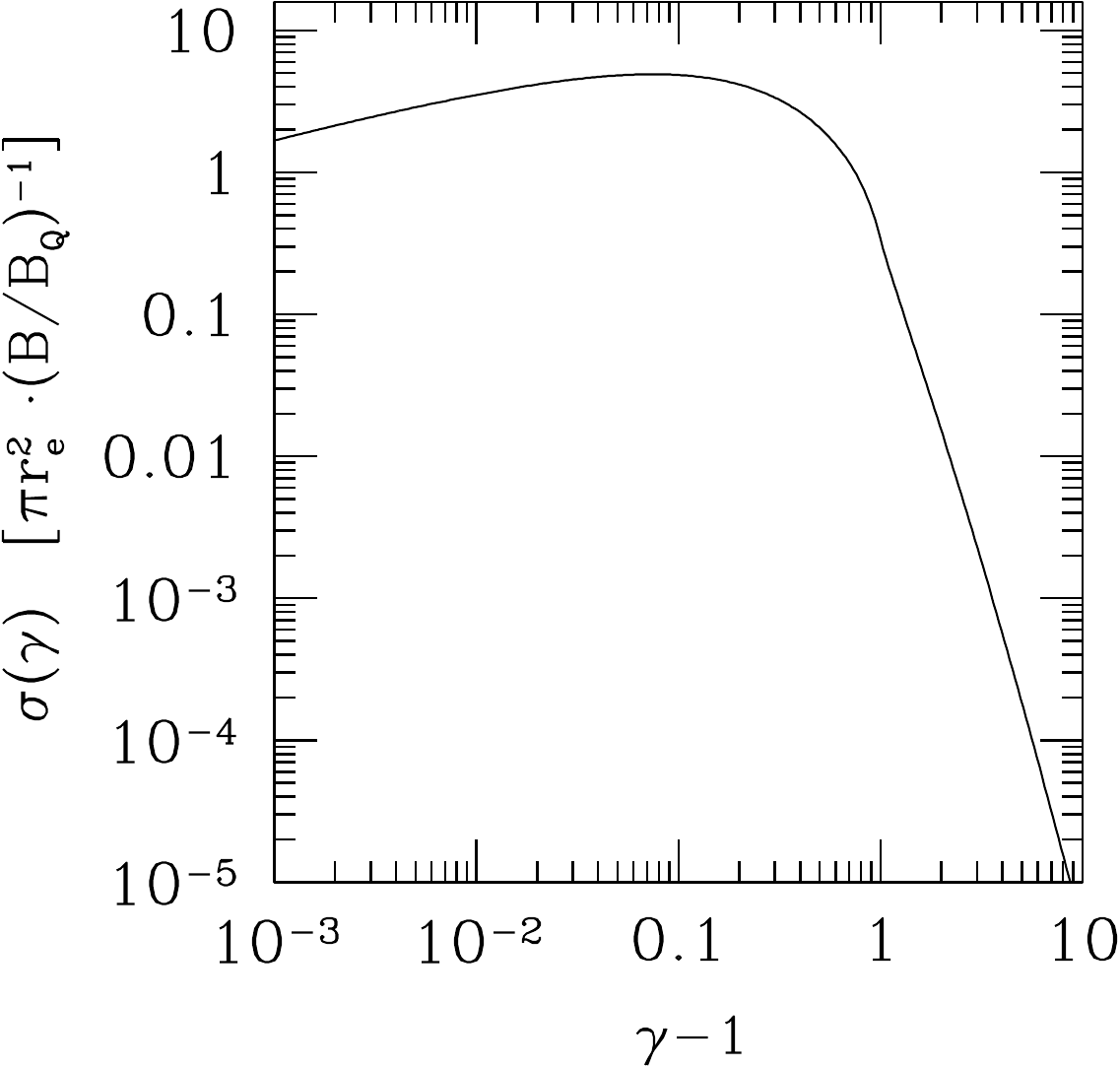}
\caption{Net cross section for two-photon annihilation, as a function of the kinetic energy of the incoming
electron and positron as measured in the center-of-momentum frame.  Only final photon states that are below the
threshold (\ref{eq:eth}) for single-photon pair creation contribute to the phase-space integral in 
Equation (\ref{eq:sigann2}).  The sharp drop at $\gamma \gtrsim 2$ represents the rapidly shrinking phase space for non-pair-converting photons.}\label{2SigmaAnnEffect}
\end{figure}
 
The evaluation of the integrals in Equation (\ref{eq:sigann2}), as a function of the Lorentz factor of the incoming electron
and positron, is shown in Figure \ref{2SigmaAnnEffect}.  The behavior at low $\gamma$ lines up qualitatively with the result of
\citet{DB1980}, who found that the annihilation cross section at rest decreases with increasing $B$.  The decrease in the
cross section beyond $\gamma-1 \sim 0.1$ is due to the growing restriction on the final photon phase space from reconversion
to a pair.  A combined analytic and numerical fit, valid up to $\gamma \simeq 2$, is
\be\label{eq:annfit}
\sigma = {8\pi r_e^2\over B/B_{\rm Q}} {\beta\over\gamma^2} \left[ \frac{5}{9} - 1.29116 \beta^2 - 0.0886 \beta^4-\ln(\beta)\left(\frac{4}{3} + 0.93634 \beta^2 + 4.53916 \beta^4 \right) \right].
\ee
(Here, the leading linear term and the leading term proportional to $\ln(\beta)$ inside the brackets are derivable
analytically in the low-$\beta$ regime.)  The sharp cutoff at higher $\gamma$ is fitted as
\be\label{eq:annfit2}
\sigma =  \frac{8\pi r_e^2}{B/B_{\rm Q}}{1\over \gamma^6}\left ( 1 + \frac{2.84}{\gamma^2} + \frac{82}{\gamma^6} \right )
\ee
for $\gamma \gtrsim 2$, with the leading term representing an analytic fit.

\section{Summary}\label{s:concl}

We have derived and analyzed the rates of several key electromagnetic processes in the presence of an intense
magnetic field:  photon-electron scattering, one- and two-photon pair creation, and two-photon annihilation.  The results are presented in a form that makes them easy to apply to the analytical and numerical study of systems such as 
the inner magnetospheres of the soft gamma repeaters and anomalous X-ray pulsars, and the collisions of strongly
magnetized neutron stars.  The basic approximation made (restricting real and virtual $e^\pm$ to the lowest Landau state)
is shown to be an excellent approximation over a significant range of magnetic field strengths, $(10-10^2)B_{\rm Q}$.

A summary of our main results follows.

1. The total cross section for electron-photon scattering rises monotonically with photon energy at fixed photon direction,
 especially for $\omega_i \gtrsim m$ (Figures \ref{NRDiffSig} and \ref{NRIntSig}).   The spike in the cross section 
resulting from a $u$-channel pole (corresponding to final photon energy $\omega_f(1-\mu_f^2) \lesssim 2m$) is quantified.

2. The dominant finite-$B$ correction to the rates of electron-photon scattering, two-photon pair annihilation, and
two-photon pair creation is identified (Equations (\ref{eq:corr}) and (\ref{eq:corr2})).

3. The rate of single-photon pair creation into the lowest Landau level is determined (Equation \ref{eq:gampair}), 
based on a detailed balance argument and the single-photon annihilation rate computed by \citet{Wunner1979} and 
\citet{DB1980}.  This controls the width of final photon states above the threshold for pair creation and helps to
determine the rate of scattering-assisted pair creation.

4. Accurate analytic approximations to the strongly peaked cross section for scattering-assisted pair creation 
(Figure \ref{ComptonAssistedIntSig}) are presented in Equations (\ref{eq:poleapprox}) and (\ref{eq:sigint}).  These results will
be a useful ingredient in future (e.g. Monte Carlo) approaches to photon-electron interactions outside magnetars, given
the challenge in accurately sampling the narrow peak in the cross section. 

5. The kinematic constraints on scattering-assisted pair creation, $\gamma + e^\pm \rightarrow e^+ + e^- + e^\pm$, are
determined (Figure \ref{ComptonPCPhase}).

6. A simplified expression for the rate of two-photon pair creation into the lowest Landau state is presented,
constructed in terms of invariant quantities involving the photon 4-momenta (Equation (\ref{eq:2gpair2})).  This
cross section is greatly enhanced compared with an unmagnetized vacuum, by a factor of $\sim B/B_{\rm Q}$ (Figure 
\ref{2SigmaPC2}).

7. The cross section for the two-photon annihilation of an $e^\pm$ pair is expressed in terms of an integral
over the cross section for two-photon pair creation (Equation (\ref{eq:sigann})).   

8. The kinematic constraints on reconversion to a pair following two-photon annihilation are derived
(Equations (\ref{eq:mu2th}) and (\ref{eq:twopairs})).  The suppression of pair annihilation is evaluated as
a function of the kinetic energy of the colliding electron and positron (Figure \ref{2SigmaAnnEffect}). 
Accurate fitting formulae to the net cross section for pair annihilation are provided (Equations (\ref{eq:annfit}) and
(\ref{eq:annfit2})).

9. The distinction between Johnson-Lippmann and Sokolov-Ternov electron/positron wave functions
is clarified:  the latter are continuously related by a Lorentz boost to the wave functions with vanishing $p_z$
(Equation (\ref{eq:psiboost})).

An evaluation of the rates of electron-positron scattering, $e^\pm$-ion scattering and relativistic $e^\pm$-ion
bremsstrahlung can be found in \cite{KT2018}.

\acknowledgements
This work was supported by the NSERC of Canada.  A.K. thanks NSERC for graduate fellowship support.

\appendix

\section{Electron scattering matrix element} \label{ComptApp}

Here, we evaluate the vertex integrals $I_1$-$I_4$ given by Equations (\ref{eq:I1}) and (\ref{eq:I2}).    Each involves a 
contraction of the matrix (\ref{eq:amatrix}) with the spinors (\ref{eq:spinors}).  We include only the lowest term 
($n_I = 0$) in the sum over Landau levels in the Green function (\ref{ComptProp}), with the outgoing electron also being
confined to the lowest Landau state.  The quantities which appear in the integrals $I_1$ and $I_3$ are
\be\label{eq:76}
\left [ u_{0,a_I}^{(-1)*} \right ]^T\gamma_0 \gamma_\mu \varepsilon_i^\mu u_{0,a_i}^{(-1)} =
\varepsilon_i^z\,{ p_{z,i}(E_I +m) + p_{z,I}(E_i+m)\over 2L^2 [E_I(E_I+m)E_i (E_i+m)]^{1/2} }\,\phi_0(x-a_i)\phi_0(x-a_I);
\ee
\be\label{eq:77}
\left [ v_{0,a_I}^{(-1)*} \right ]^T\gamma_0 \gamma_\mu \varepsilon_i^\mu u_{0,a_i}^{(-1)} =
-\varepsilon_i^z\,{ p_{z,i}p_{z,I} + (E_I + m)(E_i + m) \over 2L^2 [E_I(E_I+m)E_i (E_i +m)]^{1/2} }\, \phi_0(x-a_i)\phi_0(x-a_I).
\ee
Next, we integrate over $x$ to obtain $I_1$ and $I_3$, making use of the integral formula (\ref{eq:phiint}):
\ba\label{eq:78}
I_1 \;&=&\; \varepsilon_i^z\,\,e^{-\lambda_B^2 k_{\perp,i}^2/4}\, e^{ik_{x,i}(a_i + a_I)/2}\,
{ p_{z,i}(E_I +m) + p_{z,I}(E_i+m) \over 2L^2 [E_I(E_I+m)E_i (E_i+m)]^{1/2} } \;
 (2\pi)^2\, \delta \left( \frac{a_I-a_i}{\lambda_B^2}+k_{y,i} \right) \nn
&& \quad \times \delta(p_{z,i} + k_{z,i} -p_{z,I});\nn
\ea
\ba\label{eq:79}
I_3 &=& -\varepsilon_i^z\,\,e^{-\lambda_B^2 k_{\perp,i}^2/4}\, e^{ik_{x,i}(a_i + a_I)/2}\,
{ p_{z,i}p_{z,I} + (E_i+m)(E_I +m) \over 2L^2 [E_I(E_I+m)E_i (E_i+m)]^{1/2} } \;
 (2\pi)^2\,\delta \left( \frac{a_I-a_i}{\lambda_B^2}+k_{y,i} \right) \nn
&& \quad \times \delta(p_{z,i} + k_{z,i} + p_{z,I}).\nn
\ea
The integrals $I_2$ and $I_4$ are obtained by interchanging $E_i$, $p_{z,i}$, $a_i$, and 
$\varepsilon_i^z$ with $E_f$, $p_{z,f}$, $a_f$, and $(\varepsilon_f^z)^*$ in these expressions.   We can choose
$k_x = 0$ in the initial (but not the final) state, given the freedom in the definition of the background magnetic gauge.
Substituting these expressions for $I_1$-$I_4$ into the integral (\ref{eq:Sfi1}) gives the expression (\ref{eq:sfi1}).

\section{Erratum}

This erratum is printed separately in the Astrophysical Journal, and is included for
convenience as an Appendix in this arXiv version of the paper.  The rate (\ref{eq:gampair}) of single-photon pair creation,
$\gamma \rightarrow e^+ + e^-$ (derived in Section \ref{1gamma}), and the cross section 
(\ref{eq:2gpair}) for two-photon pair creation, $\gamma + \gamma \rightarrow e^+ + e^-$ (derived in Section 
\ref{2gammareview}), are both corrected upward by a factor of 2 as the result of an
undercounting of the phase space of the created electron and positron.  The cross section for two-photon pair annihilation
does not change, meaning that the integral relation (\ref{eq:sigrelation}) must be adjusted by a factor of ${1\over 2}$.

\subsection{Single-photon Pair Creation}

The rate for $\gamma \rightarrow e^+ + e^-$, with both electron and positron confined to the lowest Landau level,
has been obtained from the cross section for single-photon pair annihilation using a detailed balance argument.  The
right-hand side of Equation (\ref{eq:db}) must be augmented by a factor 2 to account for annihilations of
both left-moving positrons with right-moving electrons, and right-moving positrons with left-moving electrons.
Then the single-photon pair creation rate (\ref{eq:gampair}) must also be adjusted upward by a factor 2, becoming
\be\label{eq:gamrev}
\Gamma_\pm(\omega,\theta) 
= 4\alpha_{\rm em} {B\over B_{\rm Q}}  {m^4\over \omega_\perp^2(\omega_\perp^2-4m^2)^{1/2}}
 e^{-(B_{\rm Q}/2B)(\omega_\perp/m)^2} \sin\theta; \quad\quad \omega_\perp = \omega\sin\theta.
\ee

This result can be derived alternatively directly from the matrix element,
\be
\Gamma_\pm = {1\over T}\int L{dp_{z+}\over 2\pi}\int L{dp_{z-}\over 2\pi} \int L{da_+\over 2\pi \lambda_B^2}
\int L{da_-\over 2\pi \lambda_B^2} |S_{fi}|^2,
\ee
where
\be
S_{fi} = -ie\int d^4x \left[\bar\Psi^{(-1)}_-(x)\right]_{p_{z-},0,a_-} \gamma_\mu A^\mu(x)
\left[\Psi^{(+1)}_+(x)\right]_{p_{z+},0,a_+}.
\ee 
The contraction of the Dirac spinors with the polarization tensor is easy to work out when the
final-state $e^+$ and $e^-$ are both in the lowest Landau state.  Making use of the normalzation
relation (\ref{eq:phiint}) and substituting the wavefunctions from Section \ref{s:qed} gives
\be
S_{fi} = -ie {\varepsilon^z m\over E (2\omega)^{1/2} L^{7/2}} e^{-\lambda_B^2\omega^2/4} \cdot (2\pi)^3 \delta^3_{fi}(E,p_y,p_z)
\ee
for an O-mode photon propagating perpendicular to the background magnetic field ($\varepsilon^z=1$; $p_{z+} = - p_{z-}$;  $E_+ = E_- = E$).  
Then 
\be
\Gamma_\pm = {e^2 m^2\over 4\pi E^2\omega \lambda_B^2} e^{-\lambda_B^2\omega^2/2}\int {da_+\over L} 
\int_{-\infty}^\infty dp_{z+} \, \delta(2E_+ - \omega).
\ee
This agrees with Equation (\ref{eq:gamrev}) after we substitute $E = \omega/2$.   

As a result of this correction, the resonant peaks in Figure \ref{ComptonAssistedIntSig} are adjusted downward by a factor $1/\sqrt{2}$.

\subsection{Two-photon Pair Creation}

The cross section for $\gamma + \gamma \rightarrow e^+ + e^-$ has been obtained by integrating
the squared matrix element over the longitudinal momenta $p_{z,\pm}$ of the created electron and positron. 
The result in Equation (\ref{eq:2gpair}) describes the creation of a pair with longitudinal momenta
$p_{z,+} = -p_{z,-}$ of a single sign.  It must be raised by a factor of 2 to represent the total cross 
section for the creation of a pair with $p_{z,+} = -p_{z,-}$ either positive or negative.  Then
\be
\sigma   = \frac{2\pi}{\left | 1-\mu_{12} \right |} \left ( \frac{e^2}{4\pi m} \right )^2
\frac{B}{B_{\rm Q}}\frac{m^4}{\omega_1 \omega_2 E}
\frac{  |\varepsilon_1^z \varepsilon_2^z|^2}{|p_z|}   
\left |\frac{4p_z}{\omega_1\omega_2(1 - \mu_1 \mu_2 )^2 + 4\mu_1 \mu_2 p_z^2} \right |^2
\ee
in the center-of-momentum frame, and
\ba
\sigma &=& {32\pi r_e^2\over k_1\cdot k_2} {B \over B_{\rm Q}} {\sqrt{C_+-4m^2}\over C_+^{1/2}}  
          {C_{\perp 1}C_{\perp 2}C_+^2 m^6\over
          [C_{\perp1}C_{\perp2}C_+ + 4m^2(C_\times^2 - C_{\perp1}C_{\perp2})]^2}\nn
       &=& {32\pi r_e^2\over |1-\mu_{12}|} {B\over B_{\rm Q}} {\sqrt{C_+-4m^2} \over C_+^{1/2} (\omega_1\omega_2)^3}   
          {C_+^2(1-\mu_1^2)(1-\mu_2^2) m^6\over [C_+(1-\mu_1^2)(1-\mu_2^2) + 4m^2(\mu_1-\mu_2)^2]^2}
\ea
in an arbitrary Lorentz frame boosted along ${\bm B}$.  
The curves in Figures \ref{2SigmaPC1} and \ref{2SigmaPC2} must also be shifted upward by a factor 2.
These corrected expressions line up with the
$B\gg B_Q$ limit of the result given by \citet{KM1986}, which was presented as a partial cross section for
the emission of a positron and electron each with $p_z$ of a single sign.

The cross section for two-photon pair annihilation does not change.  It was obtained by direct calculation as well as
indirectly from the cross section for two-photon pair creation by the method of Section \ref{s:ann2g}.
The prefactor in the integral equation (\ref{eq:sigrelation}) must be corrected downward by a factor of ${1\over 2}$ 
to compensate for the factor of 2 increase in $\sigma_{\rm cre}$, giving
\be\label{eq:sigrelation2}
\sigma_{\rm ann} = {\lambda_B^4\over 4}\int d\mu_1 d\mu_2 \left[ {\omega_1^2\omega_2^2\over |\mu_1-\mu_2|} \cdot
|1-\mu_{12}|\sigma_{\rm cre}\right].
\ee
Note also that Equation (\ref{eq:s0sq}) for the squared matrix element was correct in the published article.

The relation (\ref{eq:sigrelation2}) above can also be obtained by a detailed balance argument.  One writes
for the reaction $e^+ + e^- \leftrightarrow \omega_1 + \omega_2$
\ba
&& (1+N_{\gamma 1})(1+N_{\gamma 2})  |\beta_+-\beta_-| {d^2\sigma_{\rm ann}\over d\mu_1d\mu_2}\Delta \mu_1 \Delta \mu_2 
\cdot 2 {dn_{e^+}\over dp_{z,+}} \Delta p_{z,+} {dn_{e^-}\over dp_{z,-}} \Delta p_{z,-} \nn
&&\quad = (1-N_+)(1-N_-)|1-\mu_{12}|\sigma_{\rm cre} \cdot
{1\over 2} {d^2n_\gamma\over d\omega_1 d\mu_1}\Delta \omega_1 \Delta\mu_1 {d^2n_\gamma\over d\omega_2 d\mu_2} 
\Delta \omega_2 \Delta\mu_2.
\ea
The factor of 2 on the left-hand side counts the two signs of $p_z$ for electrons and positrons, and
the factor of ${1\over 2}$ on the right-hand side takes into account the indistinguishability of the two photons
produced in an annihilation event.  We substitute
\ba
{dn_{e^\pm}\over dp_{z,\pm}} &=& {eB\over (2\pi)^2} N_\pm = {eB\over (2\pi)^2} {1\over e^{E_\pm/T}+1};\nn
{d^2n_\gamma\over d\omega_{1,2}d\mu_{1,2}} &=& 
{\omega_{1,2}^2\over (2\pi)^2} N_{\gamma 1,2} = {\omega_{1,2}^2\over (2\pi)^2} {1\over e^{\omega_{1,2}/T}-1}
\ea
for a thermal plasma, along with $\omega_1 + \omega_2 = E_+ + E_-$ and the Jacobian factor 
$\Delta\omega_1\Delta\omega_2/\Delta p_{z,+}\Delta p_{z,-} = |\beta_+ - \beta_-|/|\mu_1-\mu_2|$. 
The factors involving the occupation numbers $N_\pm$ and $N_{\gamma1,2}$ cancel, and we obtain
\be
{d^2\sigma_{\rm ann}\over d\mu_1d\mu_2} = 
{\lambda_B^4\over 4} \left[ {\omega_1^2\omega_2^2\over |\mu_1-\mu_2|} \cdot |1-\mu_{12}|\sigma_{\rm cre}\right].
\ee

\begin{figure}
\epsscale{0.75}
%\plotone{sigma2gPC1PDF.pdf}
\plotone{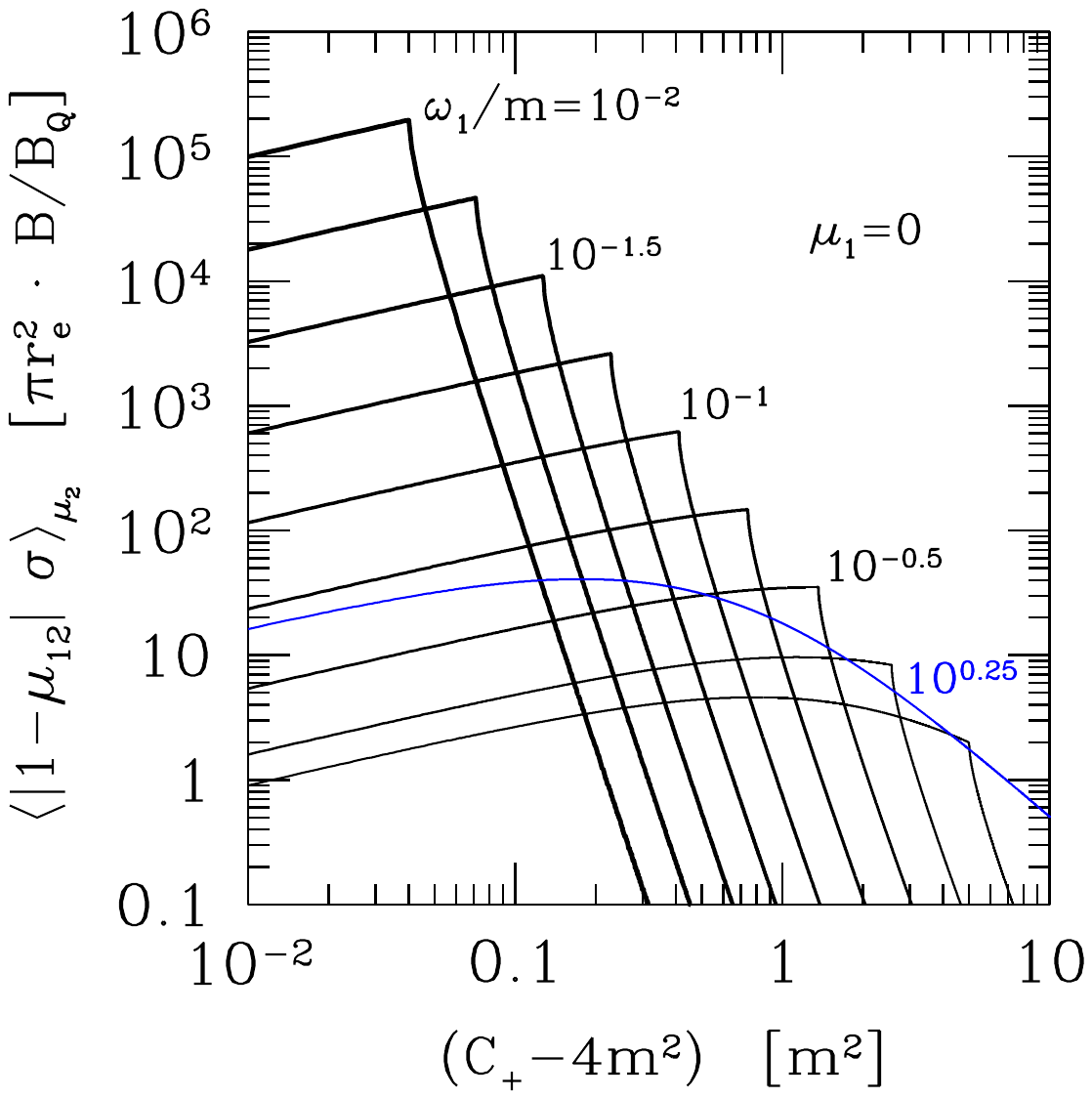}
\caption{Correction to Figure \ref{2SigmaPC1}.  The curves are raised by a factor of 2 from those 
in the published article.)  Average of the photon collision rate over direction $\mu_2$ of
the target photon, for a range of photon energy $\omega_1$ and for $\mu_1 = 0$.}\label{2SigmaPC3}
\end{figure}

\begin{figure} [h]
\epsscale{0.75}
%\plotone{sigma2gPC2PDF.pdf}
\plotone{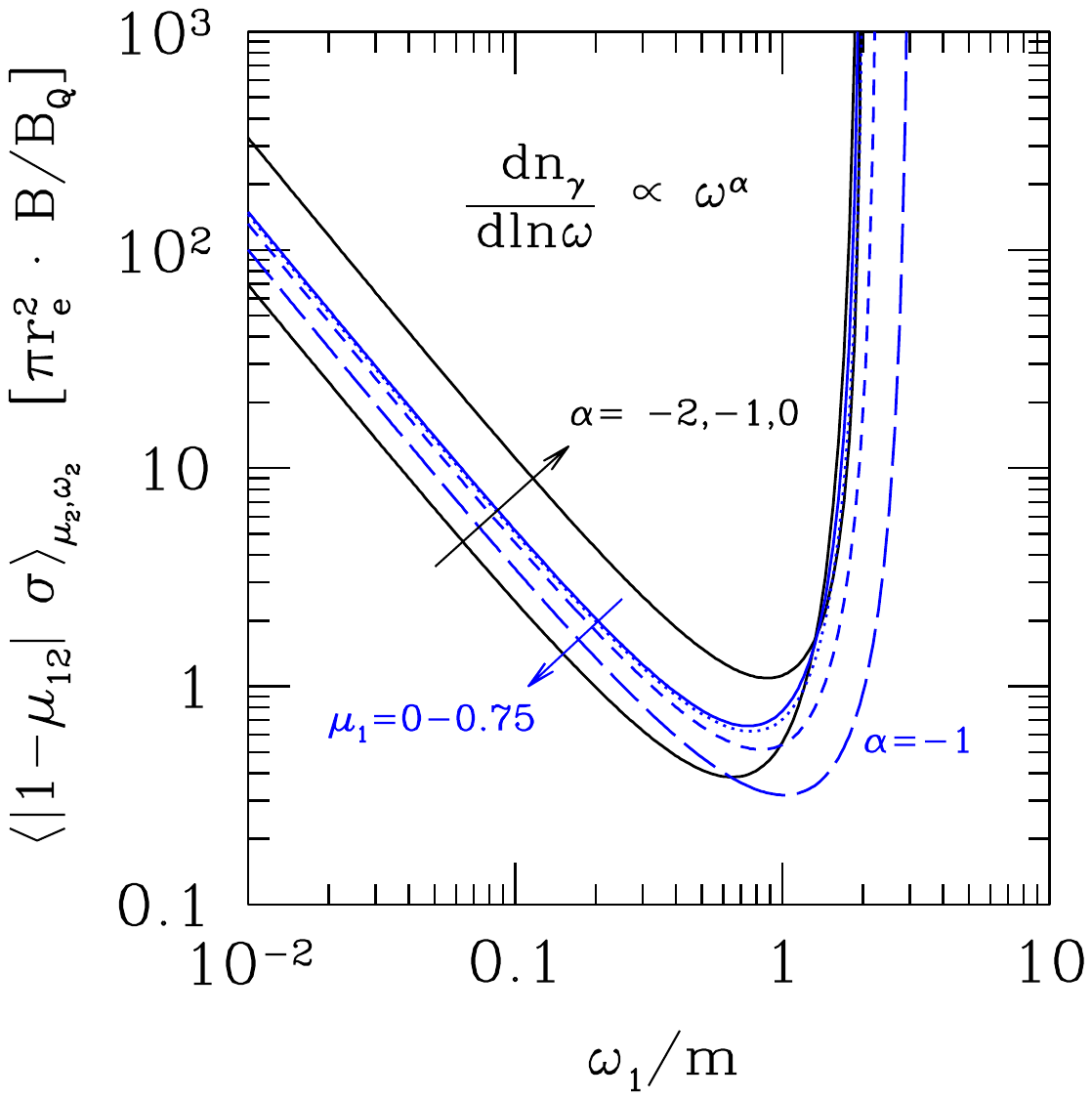}
\caption{Correction to Figure \ref{2SigmaPC2}.  The curves are raised by a factor of 2 from those
in the published article.
Average of the photon collision rate over target photon direction and energy, weighted by the frequency distribution 
of target photons, $dn/d\omega \propto \omega^{\alpha-1}$.  Black curves:  $\mu_1 = 0$ and 
$\alpha = -2$, $-1$, 0.  Blue dotted and dashed curves: range of $\mu_1$ and a flat energy spectrum 
($\alpha = -1$). Blue solid curve: $\alpha = -1$ and $\mu_1 = 0$. The rise at low frequencies and the second
rise near the threshold for single-photon pair creation both reflect the strong inverse frequency dependence of 
Equation (\ref{eq:2gpair2}), with this involving a low-frequency target photon in the latter case.}\label{2SigmaPC4}
\end{figure}

\subsection{Minor Corrections}

Labels in Figure 5:  ``Eq. (26), (30)'' should be ``Eq. (27), (31)''

Sentence before Equation (\ref{eq:2gpair2}):  ``rest-frame quantities'' should be ``center-of-momentum frame quantities.''

Sentence before Equation (\ref{eq:om12}):  ``outgoing electron and positron'' should be ``incoming electron
and positron.''

The expressions in the Appendix are constructed from electron/positron spinors using a different phase convention
than the one displayed in Equation (\ref{eq:spinors}).  As a result, Equations (\ref{eq:76}) and (\ref{eq:78}) should be
multiplied by $-1$, and Equations (\ref{eq:77}) and (\ref{eq:79}) should be multiplied by $i$, with an additional
factor of $-1$ for $I_4$.  This shift in spinor
phase convention does not affect any of the results presented in the paper.

\end{document}